\documentclass[fleqn,usenatbib]{mnras}

\usepackage{newtxtext,newtxmath}

\usepackage[T1]{fontenc}

\DeclareRobustCommand{\VAN}[3]{#2}
\let\VANthebibliography\thebibliography
\def\thebibliography{\DeclareRobustCommand{\VAN}[3]{##3}\VANthebibliography}

\usepackage{graphicx}	
\usepackage{amsmath}	
\usepackage{tabularx}
\usepackage{multirow}
\usepackage{mhchem}
\usepackage{multicol}
\usepackage{subcaption}
\usepackage{siunitx}
\usepackage{soul}
\defcitealias{Bean_2023}{BN23}
\defcitealias{gagnebin2024}{GB24}
\usepackage{makecell}

\title[Revisiting the Atmosphere of HD-149026 b]{Revisiting the Atmosphere of HD-149026 b: The Role of Stellar Abundances and Choice of Opacities in Exoplanet Atmosphere Modeling}

\author[Baghel, Goyal \& Shukla]{
Priyanka Baghel,$^{1,2}$
 Jayesh Goyal,$^{1,2}$\thanks{E-mail: jgoyal@niser.ac.in}
Gaurav Shukla$^{1,2}$
\\
$^{1}$School of Earth and Planetary Sciences (SEPS), National Institute of Science Education and Research (NISER), Jatni, Khordha-752050, Odisha, India\\
$^{2}$ Homi Bhabha National Institute, Training School Complex, Anushaktinagar, Mumbai 400094, India\\
}

\date{Accepted XXX. Received YYY; in original form ZZZ}

\pubyear{2025}

\begin{document}
\label{firstpage}
\pagerange{\pageref{firstpage}--\pageref{lastpage}}
\maketitle

\begin{abstract}

Planet formation occurs within the same molecular cloud as the host star, suggesting a link between the elemental abundances of star and the planet. Exoplanet atmosphere studies often assume solar abundances for host stars, however, specific host star abundances might lead to more accurate constraints. In this work, we perform sensitivity studies for a metal rich stellar host HD-149026 and its exoplanet HD-149026b, to understand the effect of solar versus stellar abundance choice on the $P$-$T$ profiles, equilibrium chemical abundances and emission spectra, using self-consistent atmosphere models. We find that the differences are dependent on the model parameters, particularly C/O ratio, and for HD-149026b the difference in the eclipse depth is maximum $\sim$80 ppm, for C/O between 0.75-0.85. Recent JWST NIRCam observations of HD-149026b have yielded widely varying metallicity ranges, highly super-solar (59-275$\times$) using chemical equilibrium retrievals and 12-31$\times$ solar using self-consistent models, both using solar abundances. In this work, we constrain the metallicity of HD-149026b to be 53–113$\times$ solar, with solar abundances and 39–78$\times$ stellar, with stellar abundances. We constrain the self-consistent $P$-$T$ profile of HD-149026b to be substantially cooler (upto 500\,K) than the self-consistent best-fit model in the previous work, in the emission spectra probed region, thus requiring higher CO$_2$ abundance to explain the observations, leading to comparatively higher metallicity constraint. We find that the inclusion of Fe opacity in computing self-consistent $P$-$T$ profiles for HD-149026b in our models is the major reason for these differences. We constrain the $C/O$ ratio to 0.47-0.68 and the heat redistribution factor to 0.70-0.76, indicating higher heat redistribution than previously estimated.

\end{abstract}






\section{Introduction}
Planets form from the same molecular clouds as their host stars \citep{  2018_Lammer, 2020_Liu, 2022_Curie}. Therefore, the host star elemental abundances can directly influence planetary composition, including that of the planetary atmosphere. In the initial stages of planet formation, the planetary atmosphere are expected to have a primordial hydrogen-helium dominated atmosphere, which can then evolve with time. Therefore, the measured abundance differences between the host star and the planetary atmosphere, can inform about the evolutionary stage of the planetary atmosphere as well. Therefore, obtaining robust constraints on the abundances of various species in the planetary atmosphere is crucial. 
Solar elemental abundances have typically served as initial abundance values in exoplanet atmosphere models, primarily due to non-availability of elemental abundances for planet hosting stars. However, in the recent years
various studies have focused on constraining the elemental abundances of planet hosting stars \citep{Bodaghee_2003, Mena_2018, Liu_2020,Polanski_2022}. Therefore, in this work we highlight the differences that we may obtain by assuming host star elemental abundances as compared to solar abundances, in exoplanet atmosphere models. 

HD-149026 b, a gas giant discovered in 2005 \citep{2005_Sato}, has been the subject of many studies due to its unique properties, such as its high density and significant metal content \citep{2005_Sato, 2006_Fortney, 2008_Fortney}. Initial observations using the Spitzer Space Telescope \citep{2007_Harringtone, 2009_Knutson} suggested a hot, metal-rich atmosphere with potential thermal inversions. While these early studies provided valuable insights, recent observations of HD-149026 b with the James Webb Space Telescope (JWST) allows us to place the most robust constraints. \citet{Bean_2023} (hereafter \citetalias{Bean_2023}) utilized JWST NIRCam emission spectra observations of HD-149026 b to detect water vapor (H$_2$O) and carbon dioxide (\ce{CO2}) in the atmosphere of HD-149026 b. They constrained the atmospheric metallicity to be very high (59-275 $\times$ solar) and carbon-to-oxygen ratio (C/O) to be 0.84 $\pm$ 0.03, using equilibrium chemistry retrievals. \citet{gagnebin2024} (hereafter \citetalias{gagnebin2024})  further interpreted the same JWST NIRCam observations of HD-149026 b by performing grid-retrievals using self-consistent radiative-convective-thermochemical equilibrium (RCTE), pressure-temperature ($P$-$T$) profiles.  \citetalias{gagnebin2024} constrained the metallicity to much lower value of 12 - 31 $\times$ solar metallicity and a $C/O$ ratio value of $0.67_{-0.27}^{+0.06}$. \citetalias{gagnebin2024} attributed this large difference in the metallicity constraint, to the difference between the RCTE $P$-$T$ profiles and the parametric $P$-$T$ profiles adopted in \citetalias{Bean_2023} for their equilibrium chemistry retrieval, specifically in the upper atmosphere. 

In this work we revisit the interpretation of JWST NIRCam emission spectra observation of HD-149026 b with new grids of self-consistent RCTE $P$-$T$ profiles and corresponding model emission spectra generated using ATMO, utilizing solar as well as host star elemental abundances. We report metallicity and $C/O$ ratio constraints that we obtain by assuming host star elemental abundances compared to solar abundances for HD-149026 b. We also compare our results with those of \citetalias{gagnebin2024} and \citetalias{Bean_2023}. We show a detailed comparison with \citetalias{gagnebin2024}, since we also adopt a similar grid-retrieval methodology. RCTE $P$-$T$ profiles are shaped by the opacity sources included in the model and their relative absorption cross-section in the optical and infrared wavelength regions, highlighted by many previous works \citep{2008_Fortney, Burrows2008, Hansen2008, 2015_Molliere, Goyal2020}. Therefore, in this work, we also discuss how different opacities shape the RCTE $P$-$T$ profile of HD-149026b, across a range of parameters and their values.

This paper is organized as follows. In Section \ref{sec:method} we detail the methodology of our work, followed by the sensitivity tests showing the effect of solar versus host star abundances in Section \ref{sensitivity}. In Section \ref{Results}, we show the results that we obtain for HD-149026 b, with solar and stellar abundances grids in section \ref{Sec:Solar_result} and \ref{Sec:Stellar_result}, respectively and comparison with other previous studies in Section \ref{Sec:gagnebin_result}. We finally present our discussion and conclusions in Section \ref{discussion} \& \ref{Conclusion}, respectively.

\section{Methodology}\label{sec:method}

\subsection{Atmospheric model}

In this work, we use  ATMO \citep{Tremblin_Amundsen_2015, Drummond_2016, Goyal_2018, Goyal2020}, a state-of-the-art 1D-2D planetary atmosphere model to simulate the atmosphere of HD-149026 b.
ATMO has been extensively used to model and interpret the observations of exoplanet atmospheres both as a forward model \citep[e.g.,][]{Nikolov2018, Carter2020, Goyal2021, JWSTteam_2023, Rustamkulov_2023}, as well as an inverse model \citep[e.g.,][]{Evans2018, Lewis2020, Barat2024}. In this work, we simulate the atmosphere of HD-149026 b with ATMO forward model by computing radiative-convective equilibrium $P$-$T$ profiles consistent with equilibrium chemistry for a range of model parameters. Equilibrium chemical abundances are computed using the Gibbs energy minimization method based on \citet{gordon1994computer} and detailed in \citet{Drummond_2016}, initialized by specified elemental abundances. The equilibrium chemistry scheme also includes rainout condensation as detailed in \citet{Goyal_2019}.

We include 23 elements in our model, namely, H, He, C, N, O, Na, K, Si, Ar, Ti, V, S, Cl, Mg, Al, Ca, Fe, Cr, Li, Cs, Rb, F, and P. The equilibrium chemistry calculation utilizes the elemental abundances of these species to finally compute the abundances of various gaseous and condensate species. In ATMO, the solar elemental abundances of C, N, O, P, S, K and Fe are adopted from \citet{Caffau2011} while that of He, Li, Cs, Na, Si, Ar, Ti, V, Cl, Mg, Al, Ca, Cr, Rb, F are adopted from \citet{Asplund2009}. For a particular simulation, we adjust elemental abundances based on the specified metallicity and $C/O$ ratio. To account for metallicity, we multiply the abundances of all the elements (except hydrogen and helium) by a scaling factor. To vary the $C/O$ ratio we vary O/H, while fixing the carbon abundance following \citet{2011_Moses}. Here, the $C/O$ ratio includes all elemental forms, in both gas and condensate phases.

The self-consistent $P$-$T$ profiles are computed using correlated-k opacities with random overlap method \citep{Amundsen2017} in the radiative transfer computations. Starting with an initial guess for the $P$-$T$ profile and elemental abundances (solar or stellar), ATMO iterates over the $P$-$T$  until radiative-convective equilibrium $P$-$T$ profile consistent with equilibrium chemistry is obtained. The final result is, thermochemical equilibrium abundance profile of various species and a self-consistent $P$-$T$ profile for the atmosphere. We then use this converged $P$-$T$ profile and corresponding chemical abundances to generate the planet's emission spectra. Re-circulation factor (RCF) is used to parameterize the effect of atmospheric circulation on the $P$-$T$ profile described in detail in \citet{Arora2024}.

\subsection{Host star abundances}
 Since the star and the planet form from the same molecular cloud, their abundances are interconnected and therefore the stellar atmospheric composition influences the planet's atmosphere. Therefore, in this work, we have used elemental abundances of host star HD-149026 to model the atmosphere of the exoplanet HD-149026 b and compared that with the adoption of solar elemental abundances, a more common assumption in various exoplanet atmosphere model simulations. We adopt host star HD-149026's published stellar elemental abundances from \citet{Polanski_2022}. \citet{Polanski_2022} studied a sample of 25 exoplanet host stars that were observed by JWST during its first year of operation using a data-driven machine learning technique. They published several stellar parameters for these stars, such as the effective temperature ($T_{eff}$), the projected rotational velocity (v $sin(i)$), and the surface gravity (log(g)). Additionally, the authors constrained the elemental abundances of 15 elements (C, N, O, Na, Mg, Al, Si, Ca, Ti, V, Cr, Mn, Fe, Ni, Y) in these exoplanet hosting stars.

\begin{table}
\begin{tabular}{ |p{2.5cm}|p{2.5cm}|p{2.5cm}|  }
\hline
Molecule Name & Solar Abundance relative to hydrogen (\citep{Asplund2009,Caffau2011}) & HD-149026 b Abundance (Solar) (\citep{Polanski_2022}) \\
\hline
Carbon (C)  & $3.16 \times 10^{-4} $  & 1.4791 \\
Nitrogen (N) & $7.24 \times 10^{-5} $ & 2.884 \\
Oxygen (O) & $5.75 \times 10^{-4} $ & 1.6982 \\
Sodium (Na) & $1.74 \times 10^{-6} $ & 2.5119\\
Silicon (Si) & $3.24 \times 10^{-5} $ & 1.7783\\
Titanium (Ti) & $8.91 \times 10^{-8} $ & 1.6982\\
Vanadium (V) & $8.51 \times 10^{-9} $ & 1.2883\\
Magnesium (Mg) & $3.98 \times 10^{-5} $ & 1.5136\\
Aluminium (Al) & $2.82 \times 10^{-6} $ & 1.6218\\
Calcium (Ca) & $2.19 \times 10^{-6} $ & 2.0892\\
Iron(Fe) & $3.31 \times 10^{-5} $ & 1.8621\\
Chromium (Cr) & $4.37 \times 10^{-7} $ & 1.8197\\
\hline 

\end{tabular}
\caption{The solar and host star elemental abundances used in the simulations. HD-149026 abundances are represented as a multiplicative factor of solar elemental abundances. HD-149026 elemental abundances are taken from \citet{Polanski_2022}. The abundance of each element in HD-149026 is elevated from solar elemental abundances as seen in the table.}
\label{tab:elements}
\end{table}

Table \ref{tab:elements} shows the elemental abundance values of HD-149026 with respect to sun obtained from \citet{Polanski_2022}. HD-149026 elemental abundances show a significant variation from solar abundances and some of the elemental abundances are quite high compared to solar values. For example, Nitrogen (N) is nearly three times as abundant, while Sodium (Na) and Calcium (Ca) are both more than twice as enriched. In contrast, elements like Vanadium (V) show a smaller increase, at about 1.29 times the solar value.

We have used these elemental abundances in table \ref{tab:elements} to initialize the equilibrium chemistry computations in ATMO and create a grid of self-consistent model atmospheres. Subsequently,  these model atmospheres initialized with stellar abundances are used to generate a grid of emission spectra, which is further used to interpret JWST emission spectra observations of HD-149026 b.

\subsection{HD-149026 b Grid of Model Simulations} \label{model}
We divided HD-149026 b's atmosphere into 50 layers, with logarithmically spaced pressures ranging from 100 bars at the bottom layer to $10^{-6}$ bar at the top layer. The measured stellar and planetary parameters from \citetalias{Bean_2023} were used as input to the model, similar to \citetalias{gagnebin2024}. We specifically used a planetary surface gravity (logg) of 3.2335 and a planetary radius of 0.723 $R_J$. For the host star, we used a radius of 1.454 $R_\odot$ and an orbital semi-major axis of 0.02544 AU for the planet. The closest model stellar spectra to the stellar parameters from the BT-Settl\footnotemark models \footnotetext[1]{https://phoenix.enslyon.fr/Grids/BT-Settl/AGSS2009/SPECTRA/} \citet{Allard2012, Rajpurohit2013} is used as an input to ATMO.

The metallicity as well as the Carbon to Oxygen (C/O) ratio determines the chemical composition of exoplanet atmosphere, which in turn impacts the thermal structure of the atmosphere. $C/O$ ratio can also provide clues on the formation and migration history of the exoplanet in the protoplanetary disk \citep{madhusudhan2012, oberg2011}. Additionally, recirculation factor (RCF) determines the transport of the energy from day-side to night-side \citep{Hansen2008, Arora2024, gagnebin2024}. Therefore,  in our model grid, the parameters that we vary are the $C/O$ ratio, metallicity, and the recirculation factor.

In the grid of model simulations, the RCF values used are 0.2, 0.3, 0.4, 0.5, 0.6, 0.7, 0.75, 0.8, 0.85, 0.9, 0.95, 1.0, 1.2, and 1.33. Here, the value of 1.33 implies no transport to the night-side of the exoplanet. The log(Z) (metallicity) values ranged from -1.0 to 2.6 in steps of 0.2. The $C/O$ values explored were 0.1, 0.2, 0.3, 0.35, 0.4, 0.45, 0.479, 0.5, 0.55, 0.6, 0.65, 0.7, 0.75, 0.8, 0.85, 1.0, 1.2, 1.4, 1.6, 1.8, and 2.0.

To compute grid of model atmospheres with self-consistent $P$-$T$ profiles we used 32 band correlated-k opacities for different gases specific to each grid. This is described in detail in \citet{Goyal2020}. The same set of gases are then used to generate the emission spectra grid of HD-149026 b, but with correlated-k opacities at R$\sim$1000 in wavelength. In total, we compute 5 separate grids of model atmospheres for HD-149026 b, each with different model choices detailed further in each sub-section.

\subsubsection{Grid with solar abundances varying O/H}\label{O/H}
To compute this grid of model simulations, we use solar elemental abundances as shown in Table \ref{tab:elements}. To vary the $C/O$ ratio in the grid, we vary the [O/H] abundance while keeping the [C/H] abundance constant. In this grid of model simulations opacities of \ce{H2O}, \ce{CO2}, \ce{CO}, \ce{CH4}, \ce{NH3}, \ce{Na}, \ce{K}, \ce{Li}, \ce{Rb}, \ce{Cs}, \ce{TiO}, \ce{VO}, \ce{FeH}, \ce{PH3}, \ce{H2S}, \ce{HCN}, \ce{C2H2}, \ce{SO2}, \ce{H-}, \ce{Fe},  along with \ce{H2-H2} and \ce{H2-He} collision induced absorption (CIA) were used, as detailed in \citet{Goyal2020}. To account for metallicity, we multiply the abundances
of all the elements (except hydrogen and helium) by a scaling factor.

\subsubsection{Grid with solar abundances varying C/H}\label{C/H}
\citet{Drummond_2019} highlighted, how the choice of varying [O/H] or [C/H] to vary $C/O$ ratio can lead to different equilibrium chemical abundances at different values of $C/O$ ratios.  Therefore, we also generate grid of model simulations with solar abundances by varying the [C/H] abundance to vary $C/O$ ratio, while keeping the [O/H] abundance constant. In this work, only in this grid, we vary $C/O$ ratio by varying [C/H], while in all the other grids we vary [O/H]. The set of opacities used in this grid is same as the previous grid.

\subsubsection{Grid with stellar abundance}\label{stellar}
One of the important aim of this work is to test the effect of initializing the model simulations with stellar elemental abundances instead of solar elemental abundances and therefore the obtained atmospheric constraints. 
Therefore, we generate a grid of model simulations adopting stellar elemental abundances as shown in Table \ref{tab:elements} where the stellar abundances are obtained from \citet{Polanski_2022}. The set of opacities used in this grid is the same as the previous two grids.

\subsubsection{Grid similar to \citetalias{gagnebin2024} without VO grid}\label{12}
To have one to one comparison of the interpretation of observations of HD-149026 b we also develop a grid of model simulations similar to that of \citetalias{gagnebin2024}. In this grid, we incorporate opacities due to species (\ce{H2O}, \ce{CO2}, \ce{CO}, \ce{CH4}, \ce{NH3}, \ce{Na}, \ce{K}, \ce{FeH}, \ce{PH3}, \ce{HCN}, \ce{C2H2} along with CIA opacities, similar to the list of species included by \citetalias{gagnebin2024} except N$_2$. The opacity contribution of N$_2$ is shaping the $P$-$T$ profile of hot Jupiter exoplanet atmospheres is negligible. In this grid, we use solar elemental abundances as done by \citetalias{gagnebin2024}.

\subsubsection{Grid similar to \citetalias{gagnebin2024} with VO grid}\label{13}
This grid of model simulations is exactly similar to the one described in the previous sub-section with just VO opacity added to the model while computing the grid, as it can have a substantial effect on the RCTE $P$-$T$ profiles and therefore the spectra, as highlighted by \citetalias{gagnebin2024}.

\subsection{Grid Fitting Techniques}
\subsubsection{Fitting with $\chi^2$ minimization}

The minimum \(\chi^2\) method helps find the model that best matches observed data by minimizing the \(\chi^2\) statistic. This statistic measures the difference between observed values and model predictions, adjusted for observational uncertainties. It is calculated as:
\begin{equation}
    \chi^2 = \sum_{i=1}^N \frac{(O_i - E_i)^2}{\sigma_i^2},
    \label{eq:1}
\end{equation}
where \(O_i\) is the observed value at each point \(i\), \(E_i\) is the model’s prediction, \(\sigma_i\) is the uncertainty in \(O_i\), and \(N\) is the total number of data points. A lower \(\chi^2\) value indicates a closer fit between the model and observations. In this work, our model predictions comprising a grid of simulations generated using ATMO are compared with the JWST observations using equation \ref{eq:1} to compute the \(\chi^2\)statistic and identify the best-fit model that explains the observations. Although minimizing the \(\chi^2\) can identify the best-fit model and therefore the best-fit parameters, obtaining robust uncertainties on the model parameters using this technique may not always be possible. Therefore, we also perform fitting using Bayesian Grid Retrieval technique, described in detail in the next section.

\subsubsection{Fitting with Bayesian Grid Retrieval}

Using a Bayesian grid retrieval technique, we fit the HD-149026 b JWST observations with five types of model grids, as explained in section \ref{model}. We use MultiNest's  \citep{Feroz_2009} python wrapper PyMultiNest as our Bayesian sampler. It uses nested sampling to explore the prior space of parameters and maximizes the log-likelihood in each iteration to fit the observed data to our model grid. The model grid has discrete parameters with points sampled at some spacing, and hence, we generate spectra using interpolation for the parameter values drawn by the nested sampler, which lies between the grid points in the model grid. We use the RBFInterpolator from \texttt{scipy.interpolate} \citep{2020SciPy-NMeth} for interpolation. The parameters in the grid-retrieval setup are recirculation factor(RCF), metallicity $(\log(Z))$ and $C/O$ ratio.

The radial basis function (RBF) interpolator is generally used for scattered data interpolation in multidimensional spaces and for both linear and nonlinear data. For this study, we employ a Gaussian kernel, which is known for its smoothness and suitability for capturing complex functional relationships. The spectra are interpolated as a function of the grid parameters. Forward models and the parameter values of each model from the model grid are used as input for the interpolator. This interpolated model is binned to the observed data, and the residue between this binned model and the observed data is calculated. $\chi^{2}$ is calculated using the residue and the observational data error. Log-likelihood is calculated using this $\chi^{2}$ for each interpolated model for the sampled parameter points from the prior space. This is repeated for each combination of the parameters sampled from the prior space until convergence till the highest log-likelihood value to get robust constraints. We generate the corner plot containing posterior probability distributions for the selected samples by nested sampler. Equally weighted samples are used to get the interpolated models to generate best-fit models. Median, $\pm 1$,$\pm 2\sigma$  models are computed using the numpy's percentile function and plotted. 

The RCTE profiles have different pressure grids for each model in the grid as ATMO uses optical depth rather than pressure for fixing each atmospheric layer vertically. The temperature for each $P$-$T$ profile was interpolated to a common log-spaced pressure grid using linear interpolation. RBF Interpolator with the linear kernel is used to interpolate the temperatures for the equally weighted selected samples, and these are used to plot the median $P$-$T$ along with their $\pm1\sigma$ bounds.

\section{Sensitivity Tests} \label{sensitivity}
To understand the impact of host star abundances on the planet's $P$-$T$ profile, chemical abundances, and emission spectra, across different parameter space, we performed sensitivity tests for HD-149026 b by comparing the model atmospheres initialized with stellar abundance and those with solar abundances. We perform the tests for two scenarios detailed in this section, which are as follows:

\begin{itemize}
    \item Vary $C/O$ value and keep metallicity ($\log$Z) and recirculation factor (RCF) constant ( Figure \ref{fig:Sens $C/O$}).
    \item Vary metallicity (logZ) and keep $C/O$ value and recirculation factor constant ( Figure \ref{fig:Sens logZ}).
   
\end{itemize}

\subsection{Case 1: Vary $C/O$ Value} \label{Sensitivity_C/O}

\begin{figure*}
    \centering
    \begin{subfigure}{\textwidth}
        \includegraphics[width=\textwidth]{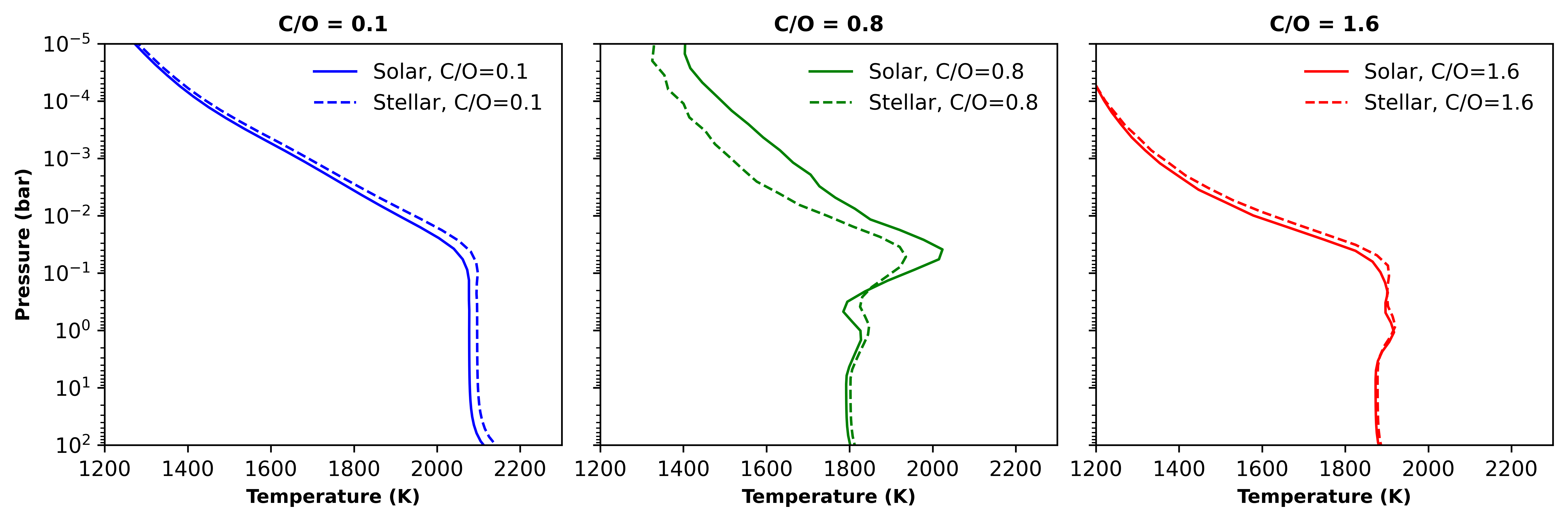}
        \caption{Comparison of solar and stellar abundances PT profiles for three different $C/O$ values while logZ and recirculation factor kept constant. The solid line represents solar abundances PT, while the dotted line represents the PT profile corresponding to stellar abundances. The blue color PT profile on the left has a $C/O$ value of 0.1, the Green PT profile in the center corresponds to a $C/O$ value of 0.8, and the Red PT profile on the right has a $C/O$ value of 1.6. The $\log Z$ value for all is 0.0 and the recirculation factor is 0.7. }
        \label{fig:Sens CO PT}
    \end{subfigure}
    
    \hfill
    \\
    \begin{subfigure}{\textwidth}
        \includegraphics[width=\textwidth]{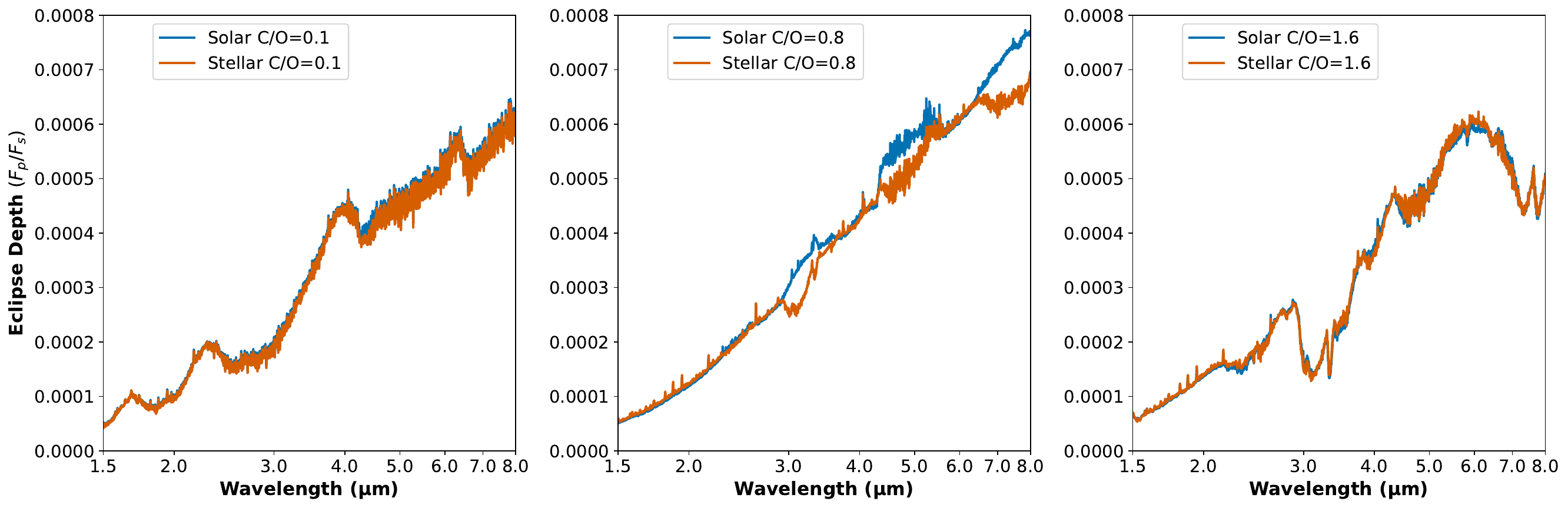}
        \caption{Comparison of emission spectra corresponding to PT profiles in \ref{fig:Sens CO PT}. The blue line represents the emission spectra obtained using solar abundances, and the orange represents the emission spectra corresponding to host star abundances.}
        \label{fig:Sens CO emission}
    \end{subfigure}
    
    \hfill
    \\
    \begin{subfigure}{\textwidth}
        \includegraphics[width=\textwidth]{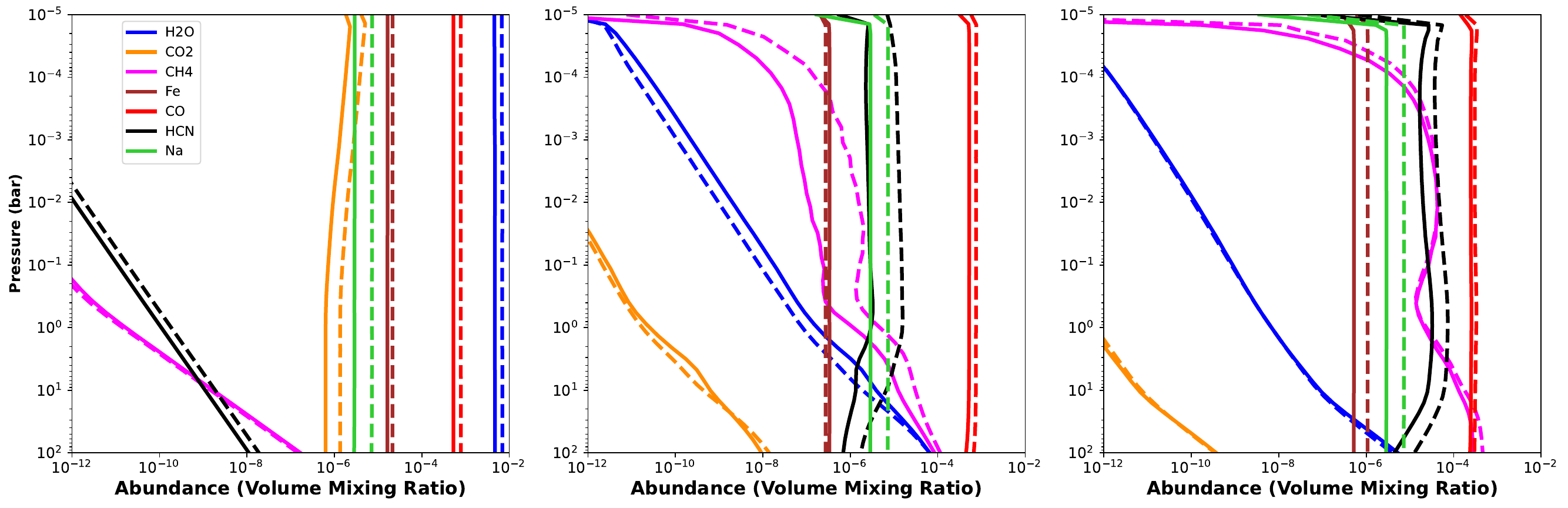}
        \caption{Comparison of a few molecules' abundances corresponding to PT profiles in fig \ref{fig:Sens CO PT} and emission spectra in fig \ref{fig:Sens CO emission}. The solid line represents molecular abundances (VMR) corresponding to solar abundance grid models, and the dashed line represents abundances corresponding to stellar abundance models.}
        \label{fig:Sens CO abun}
    \end{subfigure}
    \caption{These plots show the comparison between PT profiles, emission spectra and molecular abundances when solar and HD-149026 abundances are considered for three different $C/O$ values (0.1, 0.8, 1.6) while recirculation factor and logZ are kept constant with values 0.7 and 0.0 respectively. The overall effect of solar and host star abundances on PT profile, emission spectra and atmospheric abundances is different for different $C/O$ values. It's negligible for $C/O$ value = 0.1 and 1.6, but the difference is significant for $C/O$ value = 0.8. }
    \label{fig:Sens $C/O$}
\end{figure*}

Figure \ref{fig:Sens $C/O$} illustrates the impact of choosing solar abundances versus host star abundances on various atmospheric properties of HD-149026 b for different $C/O$ values. Specifically, Figure \ref{fig:Sens CO PT}, Figure \ref{fig:Sens CO abun} and Figure \ref{fig:Sens CO emission} show the RCTE $P$-$T$ profiles, equilibrium chemical abundances and the corresponding emission spectrum, respectively. 

These effects are examined for three different $C/O$ ratios:  
\begin{itemize}
    \item $C/O$ = 0.1 (subsolar)  
    \item $C/O$ = 0.8 (supersolar)  
    \item $C/O$ = 1.6 (supersolar)  
\end{itemize}

These values were chosen to explore a broad range of compositions, with solar $C/O$ = 0.55 as a reference. We fix \(\log Z = 0.0\) to ensure that the initial elemental abundances remain unscaled, allowing us to isolate the effect of $C/O$ variation without introducing additional metallicity-dependent changes. The RCF value of  0.7 is chosen.

It can be noticed from Figure \ref{fig:Sens CO PT} that the choice of solar versus stellar abundances leads to a notable difference in the $P$-$T$ profiles for $C/O$ = 0.8, particularly between $10^{-1}$ - $10^{-5}$ bar. However, this variation diminishes for $C/O$ = 0.1, where both abundance choices yield similar $P$-$T$ profiles due to the oxygen-dominated chemistry, and for $C/O$ = 1.6, where the atmosphere is predominantly carbon-rich regardless of the abundance choice. Similar trends are observed in the emission spectra (Figure \ref{fig:Sens CO emission}) and chemical abundance profiles (Figure \ref{fig:Sens CO abun}). For $C/O$ = 0.8, the choice of stellar abundances enhances carbon-bearing species like CH$_4$ while reducing oxygen-bearing species such as H$_2$O, resulting in spectral differences. This highlights how the choice of model initialization with different chemical abundances even with similar $C/O$ ratio can lead to different $P$-$T$ profiles, abundances and therefore the emission spectra. This will in turn result in incorrect constraints on the $C/O$ ratio of the exoplanet atmosphere from the same observational data and therefore initialization of model with stellar abundances is required to obtain more robust $C/O$ ratio constraints. 

To change the $C/O$ ratio in our model, we adjust the oxygen abundance [O/H] while keeping the carbon abundance [C/H] fixed, thus modifying the relative ratio of these elements. For the $C/O$ value of 0.8, where we see differences due to the choice of abundance sets, we observe that for the stellar case, the [O/H] value is 1.017 times higher than the solar [O/H] value i.e. it has scaled up (as the stellar [C/H] and [O/H] are higher than solar), while for the solar abundance case, [O/H] is 0.687 times solar [O/H] value i.e. it is scaled down to obtain the same $C/O$ value. On the other hand, for $C/O$ = 0.1, the oxygen abundance is enhanced by a factor of 8.13 and 5.5 for the stellar and solar abundance cases compared to solar [O/H], respectively. Although the absolute oxygen abundances differ, both remain significantly oxygen-rich compared to solar composition, leading to similar dominant molecular species and resulting in nearly identical $P$-$T$ profiles and atmospheric properties. Similarly, for $C/O$ = 1.6, the oxygen abundance is reduced by factors of 0.51 and 0.34 for the stellar and solar cases compared to solar [O/H], respectively. Since both cases have significantly lower [O/H] than solar, they exhibit similar atmospheric properties, as the dominant molecular species remain largely unchanged, again leading to identical $P$-$T$ profiles and chemical abundances.

At $C/O$ = 0.8, a distinct thermal inversion can be noticed compared to $C/O$ values of 0.1 and 1.6. To verify which molecular opacity is responsible for the inversion, we removed one species at a time from our model to generate the $P$-$T$ profile, as shown in Figure$\sim$\ref{fig:inversion_molecule_PT_sens_CtoO}. We find that this thermal inversion is caused due to \ce{Na}. Although the abundance of \ce{Na} is similar across all three $C/O$ ratios, inversion occurs only for $C/O$ of 0.8. Upon further investigation, we find that removing \ce{HCN} opacity results in much stronger inversion in the $P$-$T$ profile as shown in Figure$\sim$\ref{fig:inversion_molecule_PT_sens_CtoO} in red color, implying that inversion is suppressed by \ce{HCN} and to a lesser extent by \ce{CH4}. \ce{HCN} and \ce{CH4} due to radiative cooling, inhibit formation of inversion layer, as highlighted in \citet{2015_Molliere}. The higher abundance of \ce{HCN} and \ce{CH4} in the stellar abundance case at $C/O$= 0.8 weakens the inversion, while the lower \ce{HCN} and \ce{CH4} abundance in the solar abundance case leads to a stronger inversion. Since \ce{HCN} and \ce{CH4} abundance is high and similar for both solar and stellar cases at $C/O$ = 1.6, it effectively cools the atmosphere, preventing inversion. 

At $C/O$ = 0.1, despite the \ce{Na} abundance being similar to the other cases, no inversion is observed. Also, \ce{HCN} and \ce{CH4} abundances are very low to effectively cool down the atmosphere. In this scenario, the higher \ce{H2O} abundance leads to efficient radiative cooling, counteracting any potential heating from optical absorbers such as \ce{Na}. Since water is a strong infrared emitter, it enhances cooling in the upper atmosphere by efficiently radiating away absorbed energy. Additionally, at low $C/O$ ratios, oxygen-rich species like \ce{H2O} and \ce{CO2} dominate over carbon-bearing species like \ce{CH4} and \ce{HCN} as seen in Figure \ref{fig:Sens CO abun}. This results in a temperature-pressure profile without significant thermal inversion, as the cooling effect of \ce{H2O} outweighs the heating contributions from other absorbers.

We also observe that the differences in the emission spectra are prominent for $C/O$ = 0.8, which is due to the differences in \ce{CH4} and \ce{HCN} abundances for solar and stellar abundance case. The strong features that we see at around 3 $\mu$m and 7 $\mu$m in the emission spectra, clearly distinctive between both the cases of solar and stellar abundance are mainly due to \ce{HCN} with some contribution of \ce{CH4}. The spectral feature that we see at around 5 $\mu$m is a combined effect due to \ce{HCN} and \ce{CO}.  
The similar \ce{HCN} and \ce{CH4} abundance for both solar and stellar abundance case leads to similar spectra for $C/O$ = 0.1 and 1.6. 

Now, consider a scenario where observations of an exoplanet atmosphere produce a spectrum similar to the spectrum at  $C/O$ = 0.8 as in Figure \ref{fig:Sens CO emission}, which corresponds to stellar abundances. If a solar-abundance model is used to fit a spectrum corresponding to stellar abundances (e.g., the orange spectrum in Figure$\sim$\ref{fig:Sens CO emission} at $C/O = 0.8$), the model must compensate by increasing the abundances of \ce{HCN} and \ce{CH4} along with slight adjustments to other species. This adjustment leads to a higher inferred $C/O$ ratio than the actual value. Since $C/O$ plays a crucial role in understanding a planet’s formation history, even small changes in its value can affect the interpretation of its origin and evolution.

\subsection{Case 2: Vary logZ Value}

\begin{figure*}
    \centering
    \begin{subfigure}{\textwidth}
        \includegraphics[width=\textwidth]{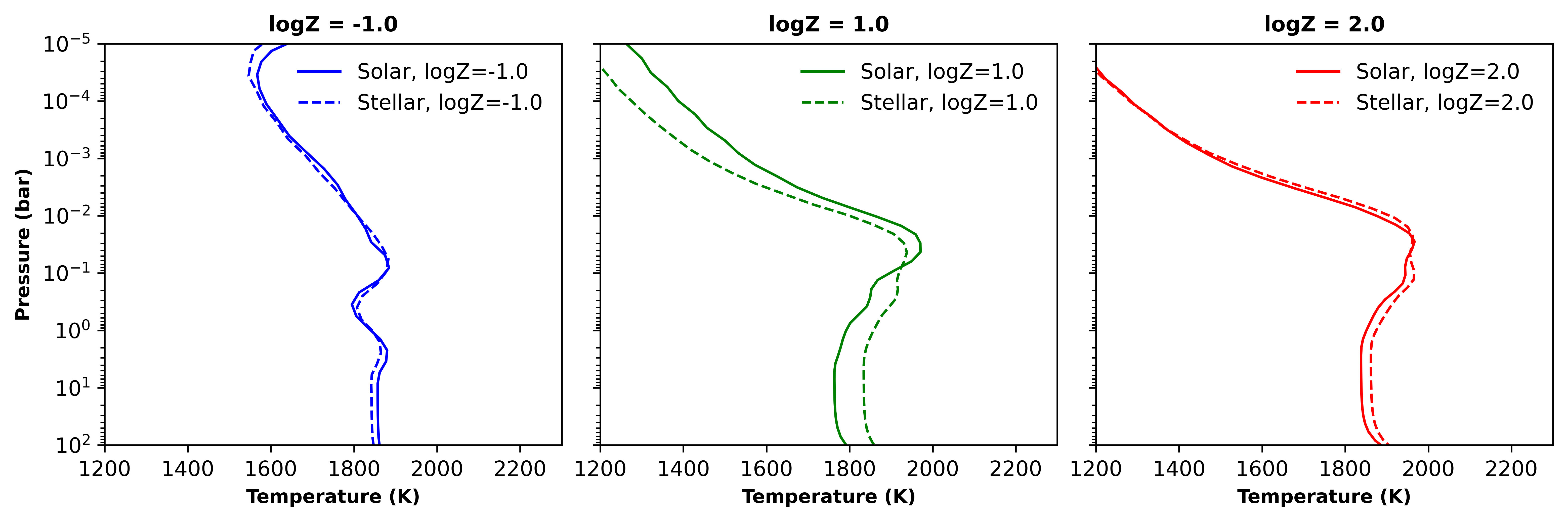}
        \caption{Comparison of HD-149026 b's PT profile corresponding to solar and stellar abundances for three different metallicity (logZ) values with constant $C/O$ and recirculation factor value. The solid line represents solar abundances PT, while the dotted line represents the PT profile corresponding to stellar abundances. The blue color PT profile on the left is for logZ value -0.1, the green PT profile in the center corresponds to logZ value 1.0, and the red PT profile on the right is for logZ value 2.0. The $C/O$ value for all the PT profiles is 0.8, and the recirculation factor value is 0.7. }
        \label{fig:Sens logz PT}
    \end{subfigure}
    
    \hfill
    \\
    \begin{subfigure}{\textwidth}
        \includegraphics[width=\textwidth]{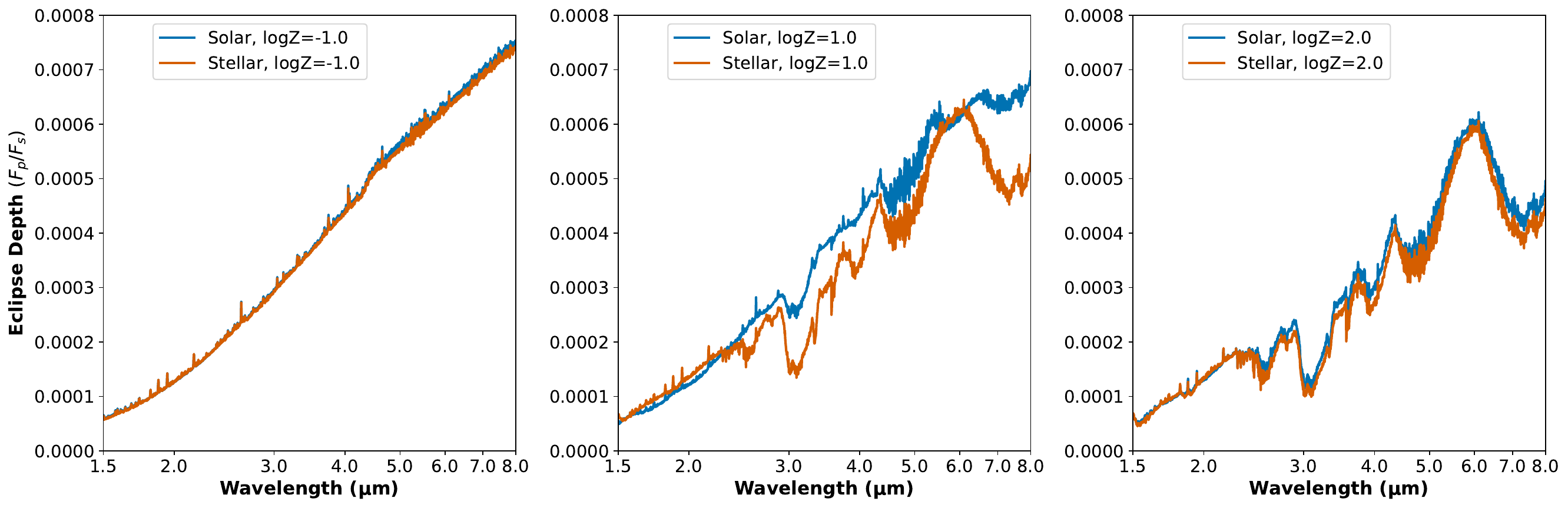}
        \caption{Comparison of emission spectra corresponding to PT profiles in fig \ref{fig:Sens logz PT}.The blue line represents the emission spectra obtained using solar abundances, and the orange represents the emission spectra corresponding to host star abundances.}
        \label{fig:Sens logz emiss}
    \end{subfigure}
    
    \hfill
    \\
    \begin{subfigure}{\textwidth}
        \includegraphics[width=\textwidth]{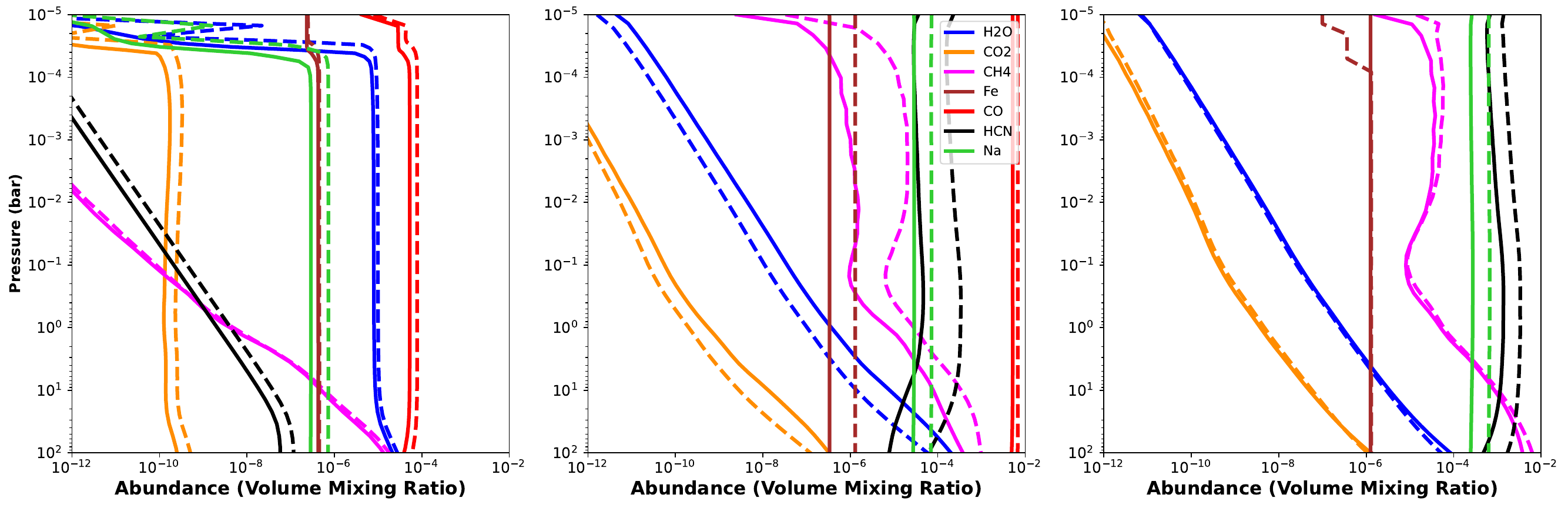}
        \caption{Comparison of a few molecules' abundances corresponding to PT profiles in fig \ref{fig:Sens logz PT} and emission spectra in fig \ref{fig:Sens logz emiss}. The solid line represents molecular abundances corresponding to solar abundance grid models, and the dashed line represents molecular abundances corresponding to stellar abundance models.}
        \label{fig:Sens logz abun}
    \end{subfigure}
    \caption{These plots show the comparison between PT profiles, emission spectra and molecular abundances when solar and HD-149026 abundances are considered for three different logZ values (-0.1, 1.0, 2.0) while recirculation factor and $C/O$ are kept constant with values 0.7 and 0.8 respectively. The overall effect of solar and host star abundances on PT profile, emission spectra and atmospheric abundances differs for different logZ values. }
    \label{fig:Sens logZ}
\end{figure*}

Figure \ref{fig:Sens logZ} illustrates the impact of selecting solar abundances versus host star abundances on various atmospheric properties for different $\log$ Z values. Specifically, Figure \ref{fig:Sens logz PT}, \ref{fig:Sens logz emiss} and \ref{fig:Sens logz abun} show the $P$-$T$ profiles, corresponding emission spectra and equilibrium chemical abundances, respectively.

These effects are examined for three different metallicity values:  
\begin{itemize}
    \item \(\log Z = -1.0\) (subsolar)  
    \item \(\log Z = 1.0\) (supersolar)  
    \item \(\log Z = 2.0\) (supersolar)  
\end{itemize}

These values were chosen to explore a broad range of metallicities, with solar metallicity (\(\log Z = 0.0\)) as a reference. We fix \(\text{C/O} = 0.8\) to explore further the differences in the atmospheric properties at this value in different metallicity conditions. It is important to note that for the stellar abundance cases the metallicity factor is multiplied to stellar abundance values. The recirculation factor is again set at RCF = 0.7.

It can be noticed from Figure \ref{fig:Sens logz PT} and \ref{fig:Sens logz abun} that the choice of solar versus stellar abundances leads to a notable difference in the $P$-$T$ profiles as well the equilibrium chemical abundances for $\log Z = 1.0$, while at $\log Z = -1.0$ and $\log Z = 2.0$ this difference is negligible. This difference in the $P$-$T$ profile and chemical abundances at $\log Z = 1.0$ leads to large differences in emission spectra with stellar and solar abundances, as seen in Figure \ref{fig:Sens logz emiss}.

We find that the choice of solar or stellar abundances does not significantly impact the atmospheric properties at very low (\(\log Z = -1.0\)) or very high (\(\log Z = 2.0\)) metallicities for HD-149026 b, as seen in Figure \ref{fig:Sens logZ}. To identify the metallicity range where the initial abundance choice differences play a significant role, we analyzed various metallicity cases. We observed noticeable differences in the atmospheric structure and emission spectra begin to emerge beyond a lower metallicity value of \(\log Z = -0.6\) and persist up to \(\log Z = 1.2\). Beyond this range, the $P$-$T$ profile, chemical abundances, and emission spectra for both solar and stellar abundance cases become nearly indistinguishable.

A possible explanation for this behavior may lie in the absolute abundance of available carbon and oxygen. For \(\log Z = -0.6\), the available carbon abundance in the stellar case is 0.3697 times solar [C/H] i.e. $1.17\times 10^{-4}$, while in the solar case, it is 0.25 times solar [C/H] ($\sim0.8\times 10^{-4}$). As metallicity decreases further, the absolute differences diminish. For example, at \(\log Z = -1.0\), the available [C/H] is 0.15 times solar [C/H] ( = $0.47\times 10^{-4}$) for the stellar case and 0.10 times solar [C/H] ($\sim0.31\times 10^{-4}$) for the solar case. For both the cases the carbon abundance is very low leading to similar chemistry and hence similar atmospheric structure. Similarly, at the upper limit of \(\log Z = 1.2\), the [C/H] abundance reaches approximately 16 times solar [C/H] ($\sim50.56\times 10^{-4}$) in the solar case and 23.44 times solar [C/H] ($\sim74.07\times 10^{-4}$) in the stellar case. Beyond this, at \(\log Z = 2.0\), the stellar [C/H] increases to 147.91 times solar [C/H] ($\sim468\times 10^{-4}$), while the solar [C/H] reaches 100 times solar [C/H] ($\sim316\times 10^{-4}$). Here also, the [C/H] value is very high for both solar and stellar abundance cases, leading to a similar kind of atmosphere. Further detailed analysis is needed to precisely describe the reasons for shaping this behaviour, which will be explored in future studies.

The differences that we find between the solar and stellar elemental abundance choices for a certain range of metallicity, can have large implications for interpreting the exoplanet atmospheric spectra and constrain their metallicity. For example, consider a case where we obtain observation for emission spectra present at centre in Figure \ref{fig:Sens logz emiss} for $logZ = 1.0$ (orange spectra), which corresponds to a model generated using host star abundances. To fit these observations with a model that assumes solar elemental abundances, the model (blue spectra for $logZ = 1.0$ in Figure $\sim$ \ref{fig:Sens logz emiss}) must adjust the corresponding temperature structure (dotted green $P$-$T$ profile in Figure$\sim$\ref{fig:Sens logZ}). Specifically, it has to decrease the temperature between 0.1 - 100 bars while increasing it in the range of 0.1 - $10^{-5}$ bars to match the stellar $P$-$T$ profile. The model will also need to enhance the abundances of carbon-bearing species like \ce{HCN} and \ce{CH4} as seen in Figure \ref{fig:Sens logz abun} in solid line to match the abundances of stellar model represented by dotted lines. It is important to note that the elemental abundance of Nitrogen in HD-149026 is 2.8$\times$ the solar value. \ce{HCN} formation requires nitrogen and to increase \ce{HCN} abundance, the model will also require high Nitrogen content. Ultimately, these adjustments will lead the model to increase the overall atmospheric metallicity to match the observed spectrum. For the same case if we assume host star abundances, we will retrieve a comparatively lower metallicity value since we use enhanced elemental abundances. 

To test this hypothesis, we used the emission spectrum with stellar abundances in Figure \ref{fig:Sens logz emiss} (orange) and generated simulated observations using PandExo \citep{2017_Batalha} for NIRCam with filters F322W2 and F444W, covering the wavelength range 2.349–5.089 $\mu$m. When we fit this simulated data with models assuming solar abundances, we indeed retrieve a model with a higher metallicity. However, the differences are very small. The best-fit metallicity using solar grid for simulated spectra with stellar abundances is 25 times solar value i.e. $log Z$ = 1.4. While when we fit these simulated observations with the stellar grid, we obtain a best-fit metallicity value of $log Z_*$= 1.0, as expected, since we simulated a model corresponding to the same metallicity value. If we consider [C/H] and [O/H] abundance only (because of varying multiplicative factors of host star abundances for each element), the multiplicative factor for stellar model will be 14 and 16 times solar respectively, while for the solar model it is 25 times solar for both [C/H] and [O/H]. Our analysis demonstrates that using solar elemental abundances instead of host star abundances can lead to an overestimation of atmospheric metallicity, even though the differences remain small in this case.

In summary, the choice of the abundance reference, whether solar or host star, can impact the interpretation of exoplanet atmosphere observations. 
However, the impact of this choice may not be uniform across all parameter space. In this work, focusing on HD-149026 b, the most pronounced differences arise within a specific range of metallicity and C/O values. This range may shift for other exoplanets and stellar hosts with different elemental compositions. It would be valuable to explore how the choice of these initial elemental abundances varies the metallicity and C/O constraints for a wide variety of exoplanets and their stellar hosts. In this section , we performed sensitivity studies with just atmosphere models. We now further use our model grids to interpret the recent JWST observations of HD-149026 b and detail the results we obtain with different model choices, including the choice of solar and stellar elemental abundances.

\section{Results} \label{Results}

We fit the JWST NIRCam observations from \citetalias{Bean_2023} using different grids of model emission spectra generated with self-consistent $P$-$T$ profiles in radiative-convective equilibrium, assuming equilibrium chemistry, computed with ATMO. These grids are generated by varying the carbon-to-oxygen (C/O) ratio, metallicity ($\log Z$), and recirculation factor, as described earlier in Section \ref{sec:method}. While varying the C/O ratio in our models, either the O/H or C/H ratio is adjusted to achieve the desired value, while all other elemental abundances are kept fixed. Similarly, when varying metallicity (log Z), the abundances of all elements are uniformly scaled by the same factor relative to solar or stellar abundances.

We create five distinct grids of models with various assumptions:
\begin{enumerate}
    \item Solar abundances (varying [O/H]) 
    \item Solar abundances (varying [C/H])
    \item Host star (HD 149026) elemental abundances from \citet{Polanski_2022} 
    \item Solar abundances with 12 molecular opacities (for comparison with \citetalias{gagnebin2024}) 
    \item Solar abundances with 13 molecular opacities, including VO (again for comparison with \citetalias{gagnebin2024})
\end{enumerate}

In this section, we present the atmospheric constraints derived for HD-149026 b using these grids and compare the results across different model assumptions.

\subsection{HD-149026 b atmospheric constraints with Solar elemental abundance grid} \label{Sec:Solar_result}

We fit JWST NIRCAM observation data from \citetalias{Bean_2023} with ATMO model grid created using solar elemental abundances, where we vary [O/H] to achieve the desired $C/O$ value. We find the best-fit model by minimizing the $\chi^2$ value and also perform grid retrievals. 

\begin{figure}
    \centering
    \includegraphics[width=\linewidth]{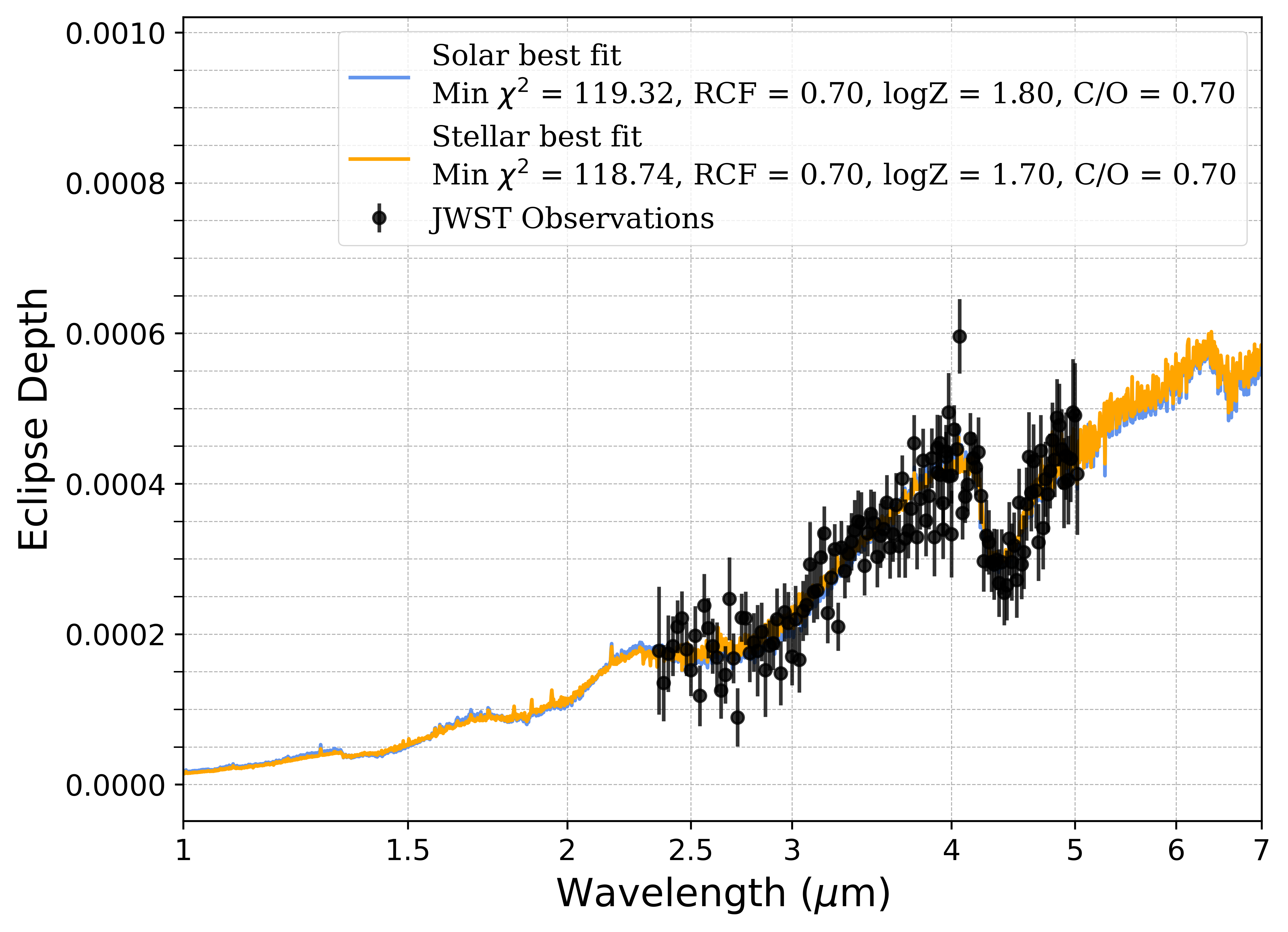}
    \caption{Best-fit emission spectra derived from a grid of ATMO models using solar elemental abundances (blue) and stellar elemental abundances (orange), fitted to JWST NIRCam observations (black). The minimum $\chi^2$ values of 119.32 and 118.74 correspond to the best-fit solar and stellar models, respectively, both at a recirculation factor of 0.7. The best-fit solar model has logZ = 1.8, while the best-fit stellar model has logZ = 1.7, both with a $C/O$ ratio of 0.7.}
    \label{fig:solar_stellar_emission_bestfit}
\end{figure}

The best-fit model that we obtain by $\chi^2$ minimization is shown in Figure \ref{fig:solar_stellar_emission_bestfit} with a recirculation factor of 0.7, $\log Z$ value of 1.8, and C/O ratio of 0.7. The minimum $\chi^2$ value for the best-fit model is 119.32, with a reduced $\chi^2$ value of 0.85 for 140 degrees of freedom (observation bins). A recirculation factor value of 0.7 indicates less than 50$\%$ of the energy due to irradiation is transported to the night side. The $\log Z$ value of 1.8 corresponds to a metallicity that is 63 times that of the Sun, indicating a significantly enhanced presence of elements heavier than hydrogen and helium in the planetary atmosphere. The $C/O$ ratio of 0.7 indicates a super-solar $C/O$ ratio. 

\begin{figure}
    \centering
    \includegraphics[width=\linewidth]{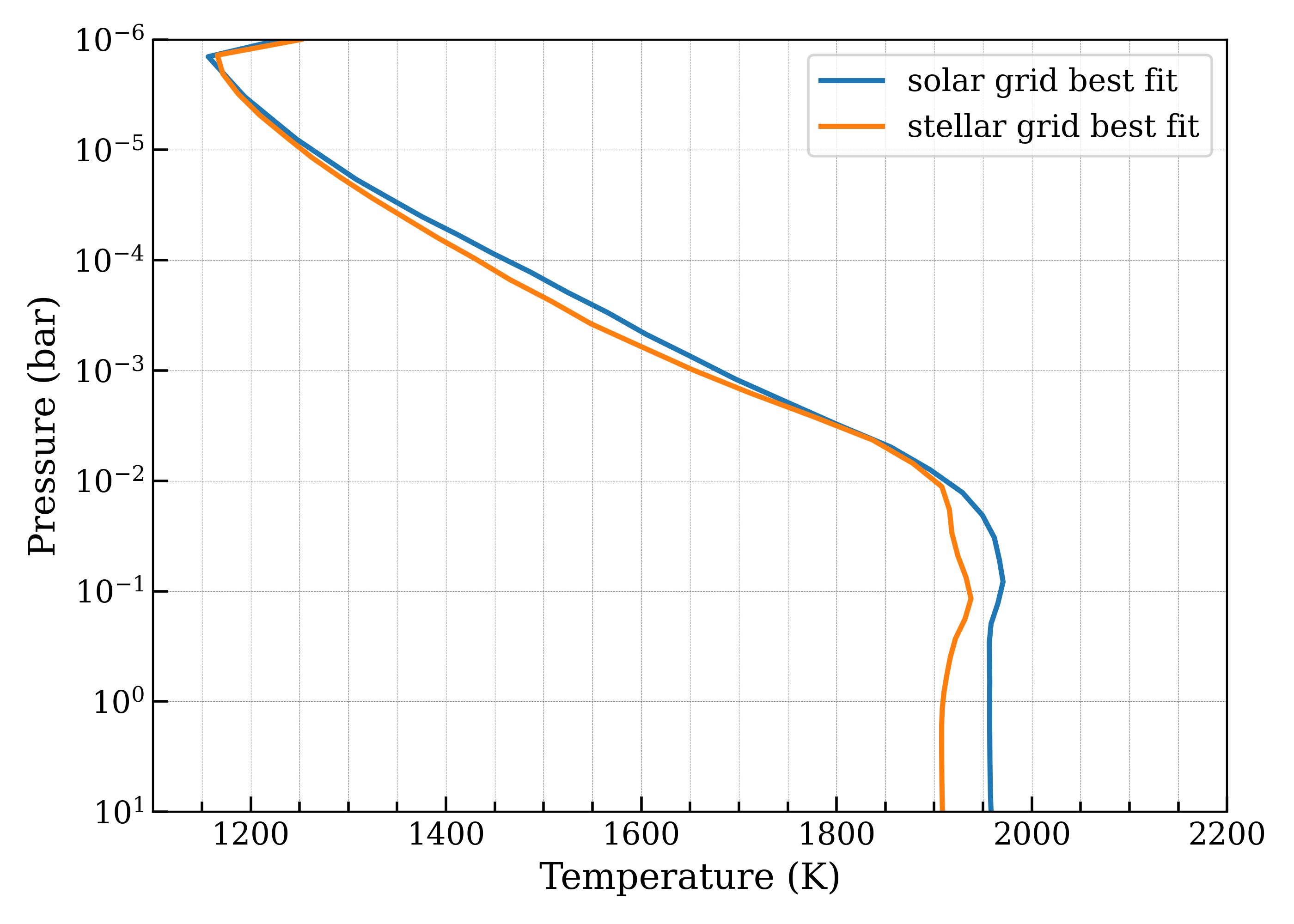}
    \caption{This Figure shows a comparison between the pressure-temperature profile for the best-fit spectra obtained from the ATMO model grid constructed using solar abundances and host star HD-149026 abundances (stellar abundances). The blue line represents the PT profile of the solar abundance best-fit model, and the orange line represents the PT profile of the stellar abundance best fit model. The Figure shows that the PT profile with host star abundances is colder than the solar abundance PT profile in the lower atmosphere.}
    \label{fig:PT_comp}
\end{figure}

\begin{figure}
    \centering
    \includegraphics[width=\linewidth]{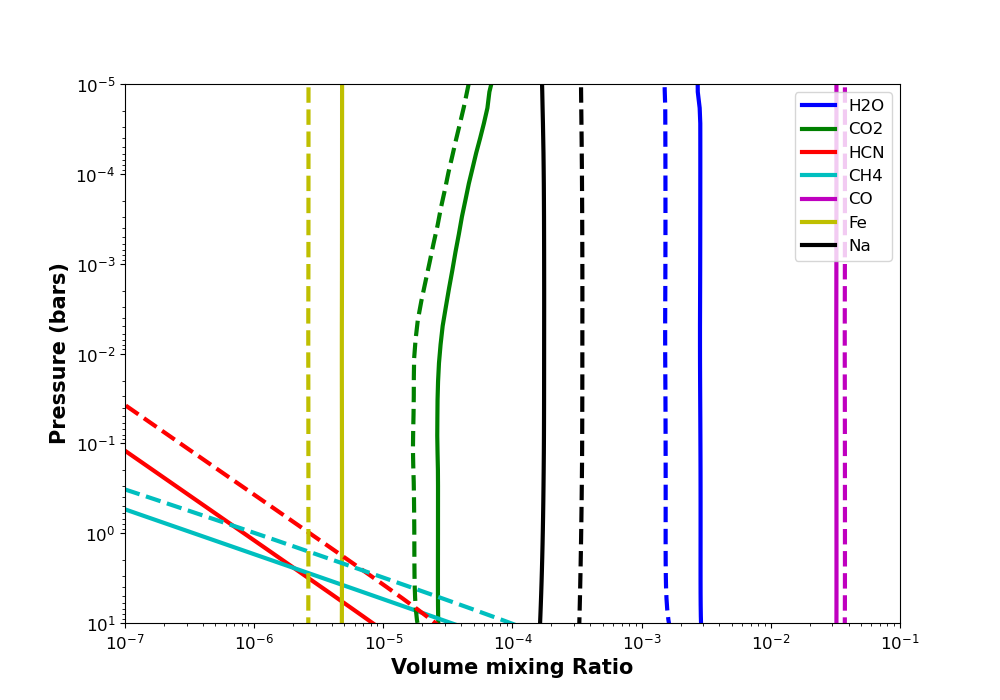}
    \caption{This plot compares the atmospheric chemical abundances of HD-149026 b corresponding to the best-fit emission spectra from solar and host star abundance grid. The solid line represents the atmospheric abundances for the solar abundance best fit, while the dotted line represents the stellar abundance best fit.}
    \label{fig:VMR_comp}
\end{figure}

The $P$-$T$ profile corresponding to the best-fit emission spectrum for the solar abundance grid is shown in Figure \ref{fig:PT_comp} (blue color). In the lower atmosphere, within the pressure range of 10-0.1 bar, the $P$-$T$ profile remains isothermal. Above this, in the 0.1-0.01 bar range, a slight temperature inversion is observed. At pressures lower than 0.001 bar, the temperature decreases with altitude as the atmosphere becomes less opaque to infrared radiation, allowing radiative cooling to dominate in these higher, less dense layers.
This overall $P$-$T$ profile structure is consistent with that expected for hot Jupiters, as shown in various previous studies \citep[for e.g][]{2008_Fortney, Guillot_2010, 2015_Molliere}.
To identify the molecular species responsible for the inversion (0.1-0.01 bar), we conducted sensitivity tests by systematically removing individual opacity sources from our model and analyzing their impact on the P-T profile. Our results indicate that \ce{Fe, Na}, and \ce{K} are the primary contributors to the temperature inversion. The volume mixing ratio of a few key molecular species from the best-fit solar abundance grid model is shown in Figure \ref{fig:VMR_comp} (Solid lines). The best-fit model shows higher abundance of dominant oxygen bearing species such as \ce{H2O} and \ce{CO2} compared to \ce{HCN} and \ce{CH4}, consistent with the best-fit $C/O$ value of 0.7. It suggests the atmosphere has more oxygen than carbon which favors the formation of \ce{CO2} and \ce{H2O} discussed in various previous works \citep[for e.g.]{2015_Molliere}. To understand the contributions of these molecules to the best-fit emission spectra, we also generate emission spectra by selectively removing the opacity of individual molecules from the model. The opacity contribution of each species to the emission spectra is shown in Figure \ref{fig:opacity}. \ce{H2O} features (blue) dominate the spectrum across all the wavelengths accept between 4 to 5 $\mu$m range, when compared to all opacity simulations (black). \ce{CO2} spectral feature dominates 4 to 4.5 $\mu$m region, followed by CO features from 4.5 to 5.5 microns. 

We also further perform grid-retrieval on the NIRCam observations using our solar abundance grid to determine the constraints on the atmospheric parameters, as well as their confidence intervals. Figure \ref{fig:solar posterior without s} shows the median model emission spectra to the observations and the one sigma bounds, obtained from the grid-retrieval samples. The corresponding posterior distribution of model parameters and their $P$-$T$ profiles are also shown. Using the solar abundance grid with varying [O/H] to vary C/O ratio and grid-retrieval we obtain a recirculation factor of $0.73_{-0.02}^{+0.03}$, a metallicity of $logZ =$ $1.94_{-0.21}^{+0.11}$ which corresponds to a metallicity of $87.10_{-33.39}^{+25.10} \times$ solar, and $C/O$ value of $0.61_{-0.14}^{+0.07}$. The results from minimum $\chi^2$ fitting and the grid retrieval technique are consistent for the recirculation factor and metallicity. However, the C/O ratio obtained from the grid retrieval is slightly lower than that derived from the best-fit model using minimum $\chi^2$. The median $P$-$T$ profiles from the grid-retrieval and its one sigma bounds are also consistent with best-fit $P$-$T$ profile obtained by $\chi^2$ minimization.

Our initial analysis with solar abundance grid was conducted using a grid of models where we vary [O/H] and keep [C/H] constant to get the desired $C/O$ value. However, it has been highlighted in many previous works that the choice of [O/H] or [C/H] to vary C/O ratio can lead to different abundances at same C/O ratio values \citep{Drummond_2019}. Therefore, we also perform grid-retrieval using the model grid when we change the [C/H] value and keep [O/H] constant to achieve the desired $C/O$ ratio, described in detail in section \ref{C/H}. With this solar abundance grid with varying C/H, we obtain a recirculation factor of $0.74_{-0.04}^{+0.03}$, a metallicity of $logZ =$ $1.96_{-0.27}^{+0.16}$ which corresponds to a metallicity of $91.20_{-42.22}^{+40.62} \times$ solar, and $C/O$ value of $0.60_{-0.29}^{+0.10}$. Our retrieval analysis on grids where either [C/H] or [O/H] is varied to achieve a desired $C/O$ ratio indicates that both approaches yield similar atmospheric constraints with HD-149026 b NIRCam observations. The best-fit parameters, recirculation factor, metallicity, and $C/O$ ratio remain within the 1$\sigma$ confidence intervals across both grids (reference figure is shown in supplementary material), suggesting that the atmospheric composition and thermal structure are not significantly impacted by the choice of C/H or O/H variation. This implies that, for the observations considered in this work and the range of models explored, the retrieval process is primarily sensitive to the resulting $C/O$ ratio rather than the specific pathway ([C/H] or [O/H] variation) used to achieve it.

\subsection{HD-149026 b atmospheric constraints with host star elemental abundance grid }
\label{Sec:Stellar_result}

In this section, we fit JWST NIRCAM observation data from \citetalias{Bean_2023} with ATMO model grid created using host star elemental abundances. The elemental abundances of HD-149026 are adopted from \citet{Polanski_2022}. We find the best-fit model by minimizing the $\chi^2$ value and also use the grid retrieval technique. 

Using minimum $\chi^2$ we obtain the best-fit model shown in Figure \ref{fig:solar_stellar_emission_bestfit} (orange), having a $\chi^2$ value of 118.74 and a reduced $\chi^2$ value of 0.85. This best-fit model corresponds to a re-circulation factor of 0.7, log Z value of 1.7, indicating a metallicity 50 times that of HD 149026 and C/O ratio of 0.7. It is important to note that metallicity here is not represented in terms of solar value, since each element in the host star varies differently with the corresponding solar elemental abundance. Table \ref{tab:Stellar metallicty retrieval} presents the multiplicative factor (metallicity) for elemental abundances corresponding to the best-fit model obtained using stellar abundance grid, expressed in terms of $\times$ solar abundance values for each element. Figure \ref{fig:PT_comp} shows the $P$-$T$ profile (orange) corresponding to the best-fit emission spectra. From Figure \ref{fig:PT_comp}, we note that HD-149026 b's atmosphere is cooler by approximately 50 K (orange) in the high-pressure range of 0.01–10 bar when considering host star abundances. The $P$-$T$ profile exhibits a minor temperature inversion around 0.1 bar, similar to that observed in the solar $P$-$T$  profile. To determine the species contributing to this inversion, we performed sensitivity tests by removing one opacity source at a time from our model and analyzing the resulting changes in the pressure–temperature profile. These tests indicate that the minor thermal inversion is most likely caused by the combined opacity contributions of Na, K and Fe. The corresponding figure illustrating this analysis is included in the supplementary material. However, since the temperature inversion is very small in this case, it is challenging to identify the exact species leading to this inversion.

The atmospheric structure between 0.01 and 10 bar remains nearly isothermal. In the upper atmosphere (P<0.001 bar), the temperature decreases significantly due to radiative cooling, consistent with the trend seen in the solar abundance profile. Further to investigate the chemical abundances, Figure \ref{fig:VMR_comp} illustrates the volume mixing ratios (VMRs) of significant molecular species, such as \ce{H2O}, \ce{CO2}, \ce{HCN}, \ce{CH4}, \ce{CO}, \ce{Fe}, and \ce{Na}. The nearly constant VMR of \ce{CO} across all pressure levels, indicates its chemical stability and dominance under thermochemical equilibrium at high temperatures \citep{Madhusudan2012}. The relatively low abundances of \ce{HCN} and \ce{CH4} are consistent with the elevated atmospheric temperatures, which favor the formation of \ce{CO} as the dominant carbon-bearing species. These molecular abundances directly influence the opacity structure of the atmosphere, with \ce{H2O} and \ce{CO2}  being primary contributors to the observed emission spectrum.

When we fit the observations with the grid of models generated using stellar abundances using the grid retrieval technique, we constrain the recirculation factor to $0.73_{-0.03}^{+0.03}$, metallicity to $logZ =$ $1.76_{-0.18}^{+0.12}$ which corresponds to a metallicity of $57.54_{-19.52}^{+18.31} \times$ stellar, and $C/O$ ratio to $0.62_{-0.13}^{+0.06}$ as shown in Figure \ref{fig:solar posterior without s} (pink color). The retrieved median $P$-$T$ also shows a slightly cooler temperature than the retrieved $P$-$T$ with solar abundance grid. This cooling is likely driven by the higher abundances of \ce{CH4} and \ce{HCN} in the stellar abundance median model (Figure \ref{fig:VMR_comp}) as previously discussed in the sensitivity studies in section \ref{Sensitivity_C/O}.

\subsection{Comparison between solar and stellar grid results}

One of the main motivations behind this work is to understand how assuming initial abundances as host star's elemental abundances, rather than those of the solar, affects our interpretation of the planet's atmosphere. In the previous two subsections, we presented detailed results for both cases. In this section, we compare our findings.

\begin{figure*}
    \centering

    \includegraphics[width=\textwidth]{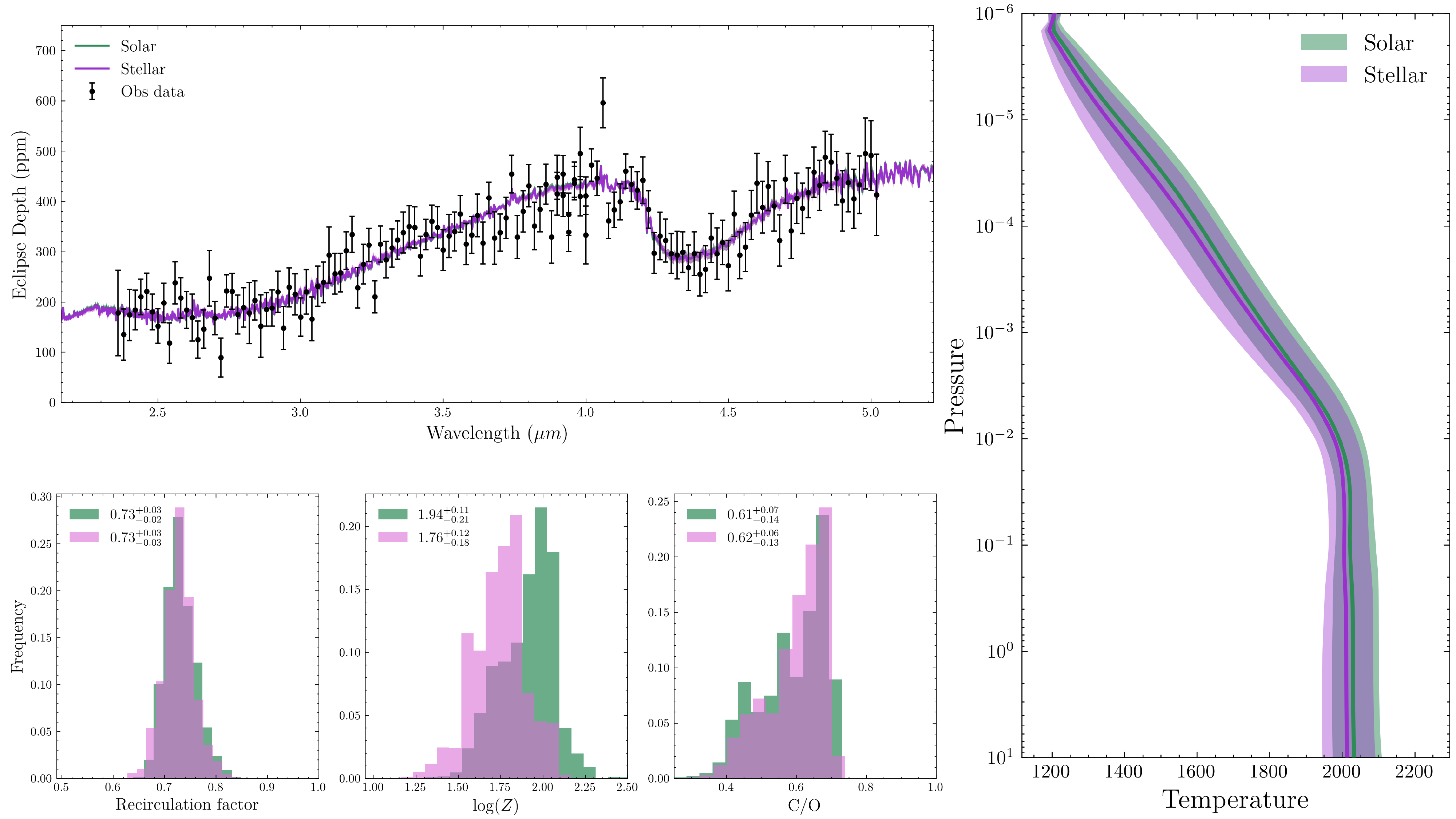}
    \caption{This plot presents the best-fit results from the solar and stellar abundance grids using grid retrieval technique. The top-left panel displays the observed data alongside the median retrieved model, with shaded regions indicating the \(1\sigma\) confidence intervals. The results from the solar abundance grid are shown in purple, while those from the stellar abundance grid are shown in green. The bottom-left panel illustrates the posterior probability distributions for the recirculation factor, metallicity (\(\log Z\)), and $C/O$ ratio from both retrievals. The titles of each posterior distribution indicate the median values along with their corresponding \(1\sigma\) credible intervals. The right panel displays the P-T profiles corresponding to the retrieved spectra from the respective grids within $1\sigma$ confidence interval. (The full posteriors are shown in supplementary material)}
    \label{fig:solar posterior without s}
\end{figure*}

Figure \ref{fig:solar posterior without s} shows the posterior probability distribution of recirculation factor, log metallicity, and $C/O$ for solar and stellar abundances models obtained using the grid retrieval technique. We find that the atmospheric metallicity of HD-149026 b is constrained to a substantially high value. The planet's atmospheric metallicity values lie in the range of 54 - 113$\times$  solar metallicity using a solar abundance grid, which is consistent with 59 - 276$\times$ solar metallicity but with lower upper bounds, than reported by \citetalias{Bean_2023} using chemical equilibrium retrievals. However, our retrieved metallicity value is significantly higher than the value reported by \citetalias{gagnebin2024}, which constrained it to 12 - 31$\times$ solar metallicity using grid retrieval. We discuss the possible reasons for these discrepancies in detail in the next section.

\begin{table}
    \centering
    \begin{tabular}{lcc}
        \hline
        Molecule & \makecell{ Abundance(log$Z$ = 1.7)\\ (relative to solar)} & \makecell{ Abundance range \\ (relative to Solar)} \\
        \hline
        Carbon (C)    & 74.13    & 57.54 - 114.82 \\
        Nitrogen (N)  & 144.54   & 112.19 - 223.87 \\
        Oxygen (O)   & 85.11     & 66.06 - 131.82 \\
        Sodium (Na)   & 125.89     & 97.71 - 194.99 \\
        Silicon (Si)   & 89.13   & 69.18 - 138.04 \\
        Titanium (Ti)  & 85.11    & 66.06 - 131.82 \\
        Vanadium (V)   & 64.57   & 50.11 - 100.00 \\
        Magnesium (Mg)  & 75.86  & 58.88 - 117.49 \\
        Aluminium (Al)   & 81.28   & 63.09 - 125.89 \\
        Calcium (Ca)  & 104.71    & 81.27 - 162.17 \\
        Iron (Fe)     & 93.33    & 72.44 - 144.55 \\
        Chromium (Cr)  & 91.20   & 70.79 - 141.25 \\
        \hline
    \end{tabular}
    
    \caption{The multiplicative factor of HD 149026 b elemental abundances relative to solar abundances for an obtained metallicity of $\log Z = 1.7$ using $\chi^2$ minimization, are shown in the second column. The abundance ranges of various elements in the atmosphere of the exoplanet HD 149026 b, expressed as a multiplicative factor of solar abundances, for a retrieved $logZ$ value of $1.76_{-0.18}^{+0.12}$ using grid retrieval, are shown in third column.}
    \label{tab:Stellar metallicty retrieval}
\end{table}

When host star abundances are considered, we retrieve a metallicity value of 38 - 76$\times$ the stellar metallicity. Unlike the solar case, we cannot express this as "$\times$ solar" because that concept assumes every element in the solar abundance is scaled by the same factor $\times$. However, in the case of stellar abundances, each element is scaled by a different factor, as shown in table \ref{tab:Stellar metallicty retrieval}. This means there’s no single common factor that applies to all elements, making it impossible to convert the metallicity into "$\times$ solar" terms. Instead, the retrieved metallicity reflects how enriched the atmosphere is compared to the host star’s composition, considering these individual scaling differences.  For example, carbon is enriched by 57.54–114.82$\times$ solar, while nitrogen ranges from 112.19–223.87$\times$ solar value.
Even though the metallicity relative to the host star is lower than that relative to the Sun, both indicate a highly enriched atmosphere. The difference arises because the host star itself is metal-rich compared to the Sun, reducing the contrast when using stellar abundances.

The recirculation factor of 0.73 obtained for both scenarios implies the distribution of less than 50$\%$ of irradiated energy from the day side to the night side of the planet. The $C/O$ value retrieved from the solar abundance grid ranges from 0.47 - 0.68, signifying more oxygen than carbon in the atmosphere of HD-149026 b with NIRCam observations. Interestingly, the $C/O$ ratio retrieved from the host star abundance grid falls within a similar range (0.48–0.68), suggesting that the atmospheric $C/O$ ratio remains consistent regardless of the choice of initial abundances. This consistency implies that the $C/O$ ratio for this particular case is less sensitive to the input abundance assumptions, likely due to the dominance of oxygen-bearing species like \ce{H2O} and \ce{CO2} in shaping the observed spectral features and similar variation of [O/H] and [C/H] in host star to that of sun. Furthermore, our sensitivity studies demonstrate that at higher or lower metallicities and C/O ratios, differences in atmospheric properties due to initial abundance assumptions tend to diminish. In the case of HD 149026 b, our findings suggest a very high metallicity for both scenarios, which could further contribute to the minimal differences observed in the retrieved parameters.

The best-fit spectra from the HD-149026 elemental abundance grid has a reduced $\chi^2$ value of 0.85, which is similar to the reduced $\chi^2$ value of best-fit spectra from the solar abundance grid (0.86). It implies that the best-fit emission spectra that we obtain for the two cases do not differ significantly in the wavelength range covered by NIRCam observations. The differences are very small and are within the error bars of observations, as shown in Figure \ref{fig:solar_stellar_emission_bestfit}.

\begin{figure*}
    \centering
    \includegraphics[width=\textwidth]{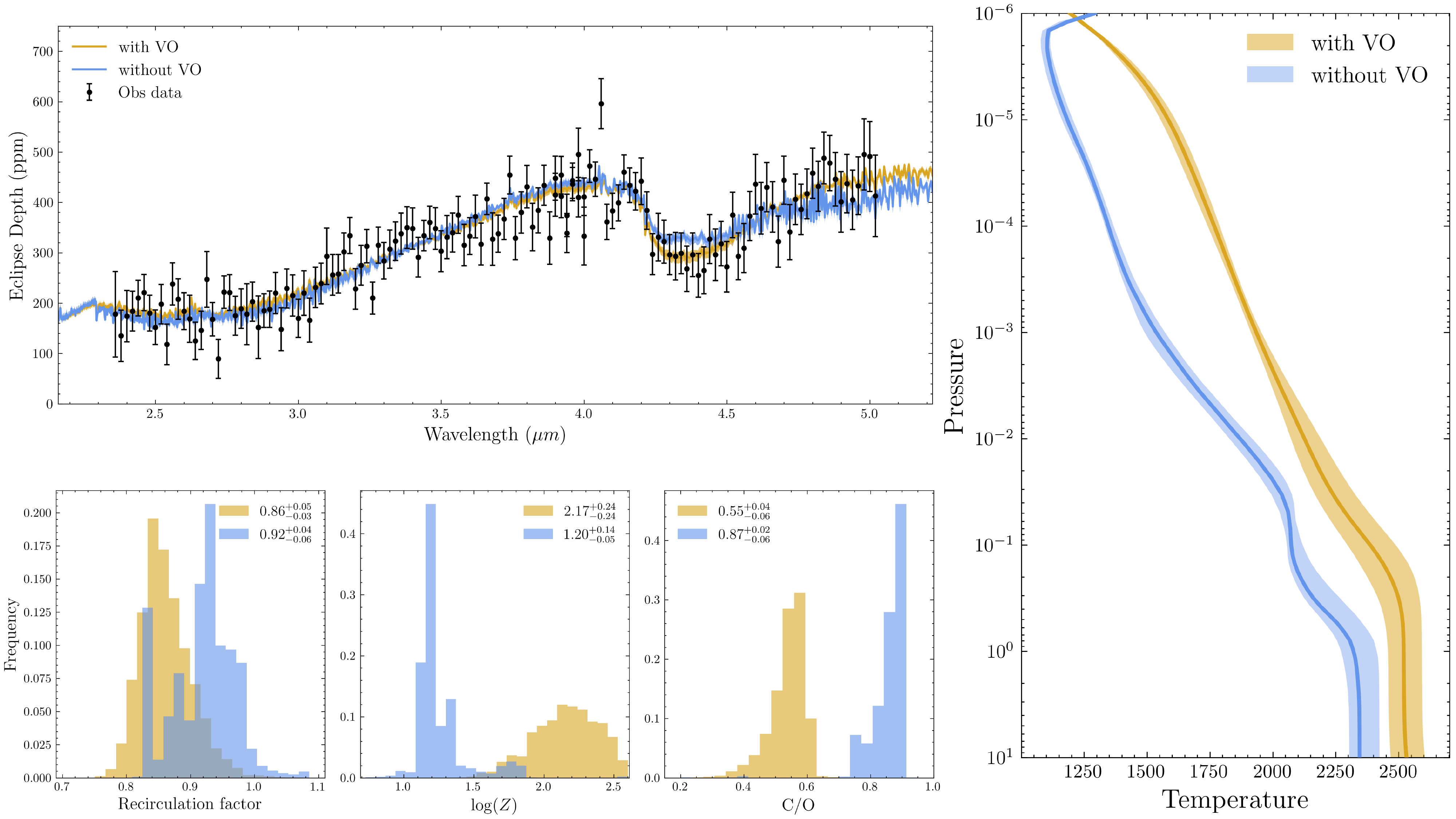}
    \caption{This plot represents the best-fit results using grid retrieval technique from grids generated using opacities of 12 molecules alone ($\sim$\ref{12}) and 12 molecules with VO ($\sim$\ref{13}) for a one-to-one comparison with \citetalias{gagnebin2024}, while assuming a constant $R_p/R_s$. The top-left panel displays the observed data alongside the median retrieved model, with shaded regions indicating the \(1\sigma\) confidence intervals. The results from the without VO grid are shown in blue, while those from the with VO grid are shown in orange. The bottom-left panel illustrates the posterior probability distributions for the recirculation factor, metallicity (\(\log Z\)), and $C/O$ ratio from both retrievals. The titles of each posterior distribution indicate the median values along with their corresponding \(1\sigma\) credible intervals. The right panel displays the P-T profiles corresponding to the retrieved spectra from the respective grids within $1\sigma$ confidence interval. (The full posteriors are shown in the supplementary material)}
    \label{fig:gagnebin_12_opacity_cons_rprs}
\end{figure*}

\subsection{Comparison of results with other works} \label{Sec:gagnebin_result}

\begin{figure*}
    \centering
    \includegraphics[width=\textwidth]{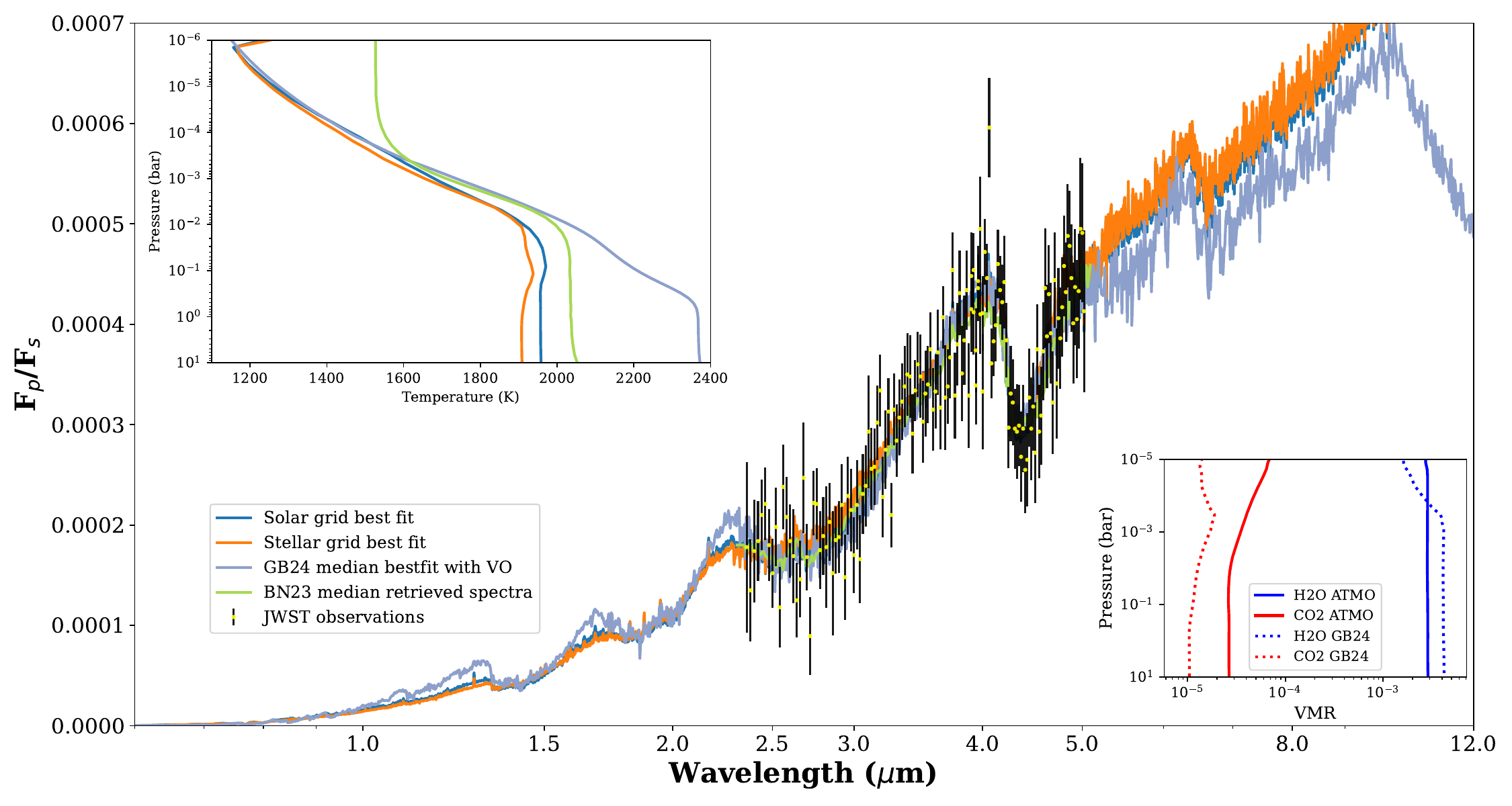}
    \caption{This plot shows a comparison of the best-fit thermal emission spectra for HD-149026 b from various studies. The blue and orange lines represent the best-fit emission spectra from this work, corresponding to the solar and stellar abundance grids, respectively. The green line shows the retrieved emission spectra from \citetalias{Bean_2023}, while the grey line represents the best-fit emission spectra from \citetalias{gagnebin2024}. The corresponding $P$-$T$ profiles are shown at the top left of the image. The \ce{H2O} and \ce{CO2} abundances for best fit $P$-$T$ profile from the solar abundance grid are shown with solid lines at the bottom right of the image along with the abundances from \citetalias{gagnebin2024} median best fit model abundances represented by dotted lines.}

    \label{fig:emiss_comp_all}
\end{figure*}

In this section, we compare our results for the atmospheric constraints on HD-149026 b with recently published studies, \citetalias{Bean_2023} and \citetalias{gagnebin2024}. \citetalias{Bean_2023} used the JWST NIRCam thermal emission spectrum to retrieve the atmospheric properties of HD-149026 b and constrained its metallicity to quite high, ranging from 
$59 - 276$ times solar metallicity with a C/O ratio value ranges between 0.81-0.87. The atmospheric constraints by \citetalias{Bean_2023} were obtained using chemical equilibrium retrieval assuming parametric $P$-$T$ profiles. Their results challenged standard planet formation models that expect lower heavy element enrichment in gas giants of the size of HD-149026 b \citep{Thorngren_2016}. The planet also has a large solid core, around 67 Earth masses as reported by \citet{2005_Sato}, indicating it may have formed through significant solid accretion, unlike typical gas giants. Later, \citetalias{gagnebin2024} interpreted the same observational data from \citetalias{Bean_2023} with a different approach. They created a grid of self-consistent model atmopsheres for HD-149026 b using PICASO \citep{Mukherjee_2023} and then fitted the observational data with the grid using a Bayesian sampler. They constrained the metallicity to 12-31$\times$ solar value, which is much lower than the value retrieved by \citetalias{Bean_2023}. \citetalias{gagnebin2024} constrained the $C/O$ ratio to $0.67_{-0.27}^{+0.06}$. They concluded that the main reason behind the difference in metallicity estimate is the difference in the behavior of the $P$-$T$ profile obtained using the self-consistent radiative-convective equilibrium model and the parametric $P$-$T$ profile assumed in the retrieval method used by \citetalias{Bean_2023}.
However, using the similar method as \citetalias{gagnebin2024} but with a different atmospheric model and model choices, we obtain a metallicity value of 54 - 115 times solar using solar abundances, which is higher than that reported by \citetalias{gagnebin2024} and lies in the range that is obtained by \citetalias{Bean_2023}.

To understand the reason for the differences in the constrained metallicity, we performed few tests. Our primary self-consistent atmosphere model grids with solar as well as stellar abundances utilized opacities for 21 molecular and atomic species as listed in section \ref{O/H}. This choice of 21 opacity species in our work was agnostic to the choice of opacities by \citetalias{gagnebin2024} and was based on previous work on self-consistent atmosphere model grids for hot Jupiter exoplanets \citep[for e.g][]{Goyal2020}. \citetalias{gagnebin2024} utilized opacities of 13 molecular and atomic species listed in 
Section \ref{12}.

We reduced our opacity molecules to 12 excluding N$_2$ as its contribution is negligible and used the median metallicity, $C/O$ value and recirculation factor values constrained by them for HD-149026 b. After making similar choices in our model, we were able to reproduce $P$-$T$ profile similar to them using ATMO as shown in Figure \ref{fig: no VO gagnebin PT}. This motivated us to create a grid of models with choice of opacities similar to that of \citetalias{gagnebin2024} for accurate comparison. Additionally, similar to their study, we created two grids: one without VO opacity as discussed earlier in this section and in Section \ref{12} and one with VO opacity included. as detailed in Section \ref{13}.

Using the grid retrieval technique with the ATMO grid similar to \citetalias{gagnebin2024} that did not include VO opacity, we obtain a metallicity value that closely matched those of \citetalias{gagnebin2024}, as shown in Figure \ref{fig:gagnebin_12_opacity_cons_rprs} (blue color). They retrieved the metallicity value to be $1.16_{-0.37}^{+0.26}$, and we found a similar value of $1.20_{-0.05}^{+0.14}$. They reported a recirculation factor of $1.17_{-0.10}^{+0.13}$, while our retrieval yielded $0.92_{-0.06}^{+0.04}$. We have a fine grid of $\sim$ 6000 models, while their grid has 2160 model atmospheres which can lead to a difference in the constraints on the recirculation factor value. Our retrieved $C/O$ value, $0.87_{-0.06}^{+0.02}$ is higher than $0.673_{-0.265}^{+0.06}$ reported by \citetalias{gagnebin2024}. A potential reason for this discrepancy is the difference in the assumed solar $C/O$ ratio: we adopted solar $C/O$ = 0.55, whereas \citetalias{gagnebin2024} used solar $C/O$ = 0.458. One more possible reason for this higher $C/O$ value could be the inclusion of rainout condensation in our models. Rainout condensation affects atmospheric chemistry by depleting certain condensable species from the gas phase.

Interestingly, the metallicity value using the ATMO grid with 12 opacities similar to \citetalias{gagnebin2024} was consistent with that of \citetalias{gagnebin2024} when we use a constant $R_p/R_s$ value in our models. When we allowed $R_p/R_s$ to vary, we observed an enhanced metallicity ($1.63_{-0.34}^{+0.31}$), a lower recirculation factor ($0.79_{-0.04}^{+0.10}$) and a $C/O$ value of $0.77_{-0.02}^{+0.02}$. The choice of constant vs. variable $R_p/R_s$ is critical, as a variable radius ratio accounts for wavelength-dependent variations due to atmospheric composition and pressure differences, which can significantly impact the obtained results. This suggests that assuming a constant $R_p/R_s$ may underestimate atmospheric complexity and lead to biases in inferred properties as discussed in detail in \citet{Taylor_2022} and \citet{2019_Fortney}.

When we performed grid retrieval using the ATMO grid with 12 opacities similar to \citetalias{gagnebin2024} in addition to VO, assuming constant $R_p/R_s$, our results diverged significantly from those of \citetalias{gagnebin2024} for metallicity as shown in Figure \ref{fig:gagnebin_12_opacity_cons_rprs}. They retrieved log metallicity value of $1.31_{-0.23}^{+0.18}$, and we find a higher log metallicity of $2.17_{-0.24}^{+0.24}$. However, the recirculation factor and C/O ratio that we constrain is consistent with that of \citetalias{gagnebin2024} as shown in Table \ref{tab:retrieval_comparison}. The reasons behind the discrepancies in the metallicity constraints in our work and that of \citetalias{gagnebin2024} with the grid that includes VO are not completely clear. As we do not have access to the VO opacity used in \citetalias{gagnebin2024}, we compared 
the VO absorption cross-sections used in ATMO to those generated 
using Helios-K \citep{2021_Grimm} using the same line-list source \citep{2016_McKemmish}. Figure \ref{fig: VO abs} in appendix shows this comparison. It can be noted that the ATMO VO absorption cross-sections are in good agreement with those generated using Helios-K. 

In addition, \citetalias{gagnebin2024} noted that gaseous TiO opacity can induce a large temperature inversion in the atmosphere of HD-1249026b. Therefore, they do not include \ce{TiO} opacity in their models. While we include TiO in both our grids (grids described in Section \ref{O/H} and \ref{stellar}) constructed considering solar and stellar elemental abundances including all the 21 opacity species, we do not observe significant temperature inversions in our results, suggesting that the presence of a strong inversion in the $P$-$T$ profile may be sensitive to particular model choices/assumptions as shown in Figure \ref{fig:rainout}. A temperature inversion in the $P$-$T$ profile is observed only when rainout condensation is excluded. When the model includes rainout condensation, no temperature inversion occurs due to \ce{TiO}.

\citetalias{gagnebin2024} suggested that the discrepancy between their lower metallicity estimates (10–20× solar) and the higher values retrieved by \citetalias{Bean_2023} (59–276× solar) likely arises from differences in the upper atmosphere (lower pressures) $P$-$T$ profiles. While \citetalias{gagnebin2024} used self-consistent $P$-$T$ profiles, \citetalias{Bean_2023} used parametric $P$-$T$ profiles in their chemical equilibrium retrievals. According to \citetalias{gagnebin2024}, the isothermal $P$-$T$ profile as constrained by \citetalias{Bean_2023} at pressures below 10 mbar requires enhanced \ce{CO2} abundance to reproduce the observed \ce{CO2} feature depth. Since \ce{CO2} abundance increases strongly with metallicity, even a slight enhancement can result in a sharp rise in the inferred atmospheric metallicity as in \citetalias{Bean_2023}. Therefore, \citetalias{gagnebin2024} concludes that since the self-consistent $P$-$T$ profile retrieved in their work is non-isothermal it can fit the observed CO$_2$ feature even with lower metallicity.

\begin{figure}
    \centering
    \includegraphics[width=\linewidth]{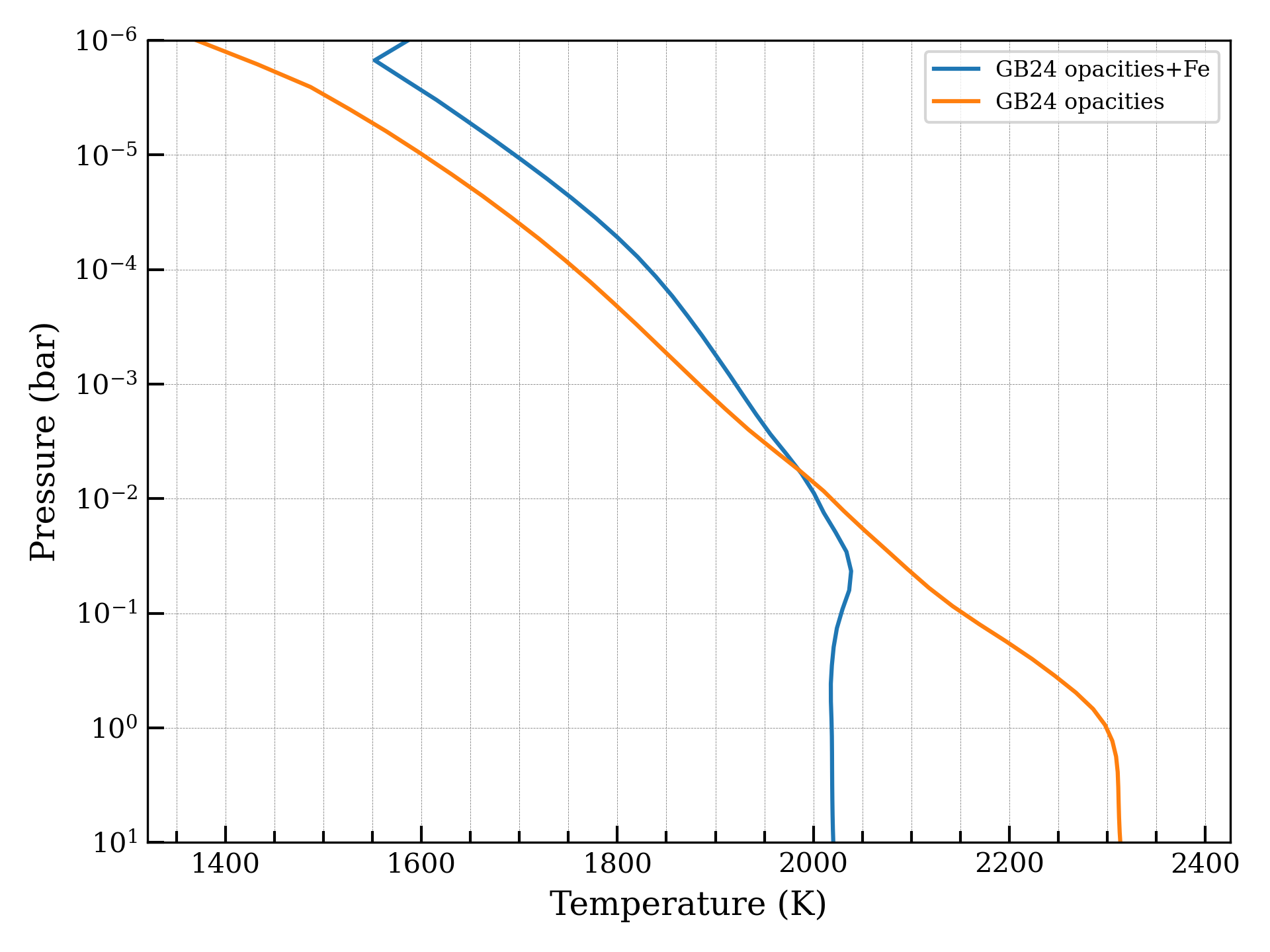}
    \caption{PT profile generated with ATMO using opacities of species used in \citetalias{gagnebin2024} including VO, along with the median retrieved values for metallicity, C/O ratio, and recirculation factor (orange). Adding Fe opacities to this model results in lower temperatures - up to 300 K cooler in the deep atmosphere, as shown in blue.}
    \label{Fe_effect}
\end{figure}

In this work, we also adopt self-consistent $P$-$T$ profiles and our best-fit $P$-$T$ profile aligns with the upper atmosphere $P$-$T$ profile reported by \citetalias{gagnebin2024} and does not match with \citetalias{Bean_2023}, shown in Figure \ref{fig:emiss_comp_all}. Despite this agreement, we obtained a higher metallicity value, which is consistent with the results of \citetalias{Bean_2023}. This finding challenges the suggestion that the discrepancy in metallicity arises solely from differences in the upper atmosphere $P$-$T$ profiles between retrievals and self-consistent models. The primary difference between best-fit $P$-$T$ profile in this work and that retrieved by \citetalias{gagnebin2024} lies in the deeper atmospheres in the pressure range of $10^{-3}-10$ bar, where the differences in our $P$-$T$ to that of \citetalias{Bean_2023} are only $\sim$50K. After a series of elaborate tests where we systematically remove one opacity at a time while computing the self-consistent $P$-$T$ profiles, we found that the difference between the best-fit $P$-$T$ in this work and that retrieved in \citetalias{gagnebin2024} is primarily due to the inclusion of \ce{Fe} opacity in our models. This addition of Fe opacity in our model lowers the temperature in the lower atmosphere (0.01 - 10 bar) by $\sim$300 K as shown in Figure \ref{Fe_effect}. The effect of Fe opacity on pressure–temperature (P–T) profiles has been previously discussed in detail by \citet{Goyal2020}. We also show the absorption cross section of Fe in Figure \ref{fig:xsec} in Appendix comparing it with other species. We can see that the absorption cross-section of Fe is comparable in magnitude to those of other strong optical absorbers such as TiO, VO, Na, and K, and in some wavelength regions even higher. This highlights the potential role of Fe in influencing the thermal structure of hot Jupiter atmospheres.
Fe in the gas phase is expected in the atmosphere of HD-149026 b with substantial abundance as shown in Figure \ref{fig:VMR_comp}. 

Moreover, since the retrieved $P$-$T$ profile in \citetalias{gagnebin2024} is hotter than that retrieved in this work, it could have Fe in the gas phase. Therefore, exclusion of Fe opacity while generating self-consistent $P$-$T$ profiles may not be completely accurate. Fe (Neutral Iron) has been marginally detected in the atmosphere of HD-149026 b by \citet{Ishizuka_2021}, further validating its inclusion while computing self-consistent $P$-$T$ profiles. 

\begin{figure}
    \centering
    \includegraphics[width=\linewidth]{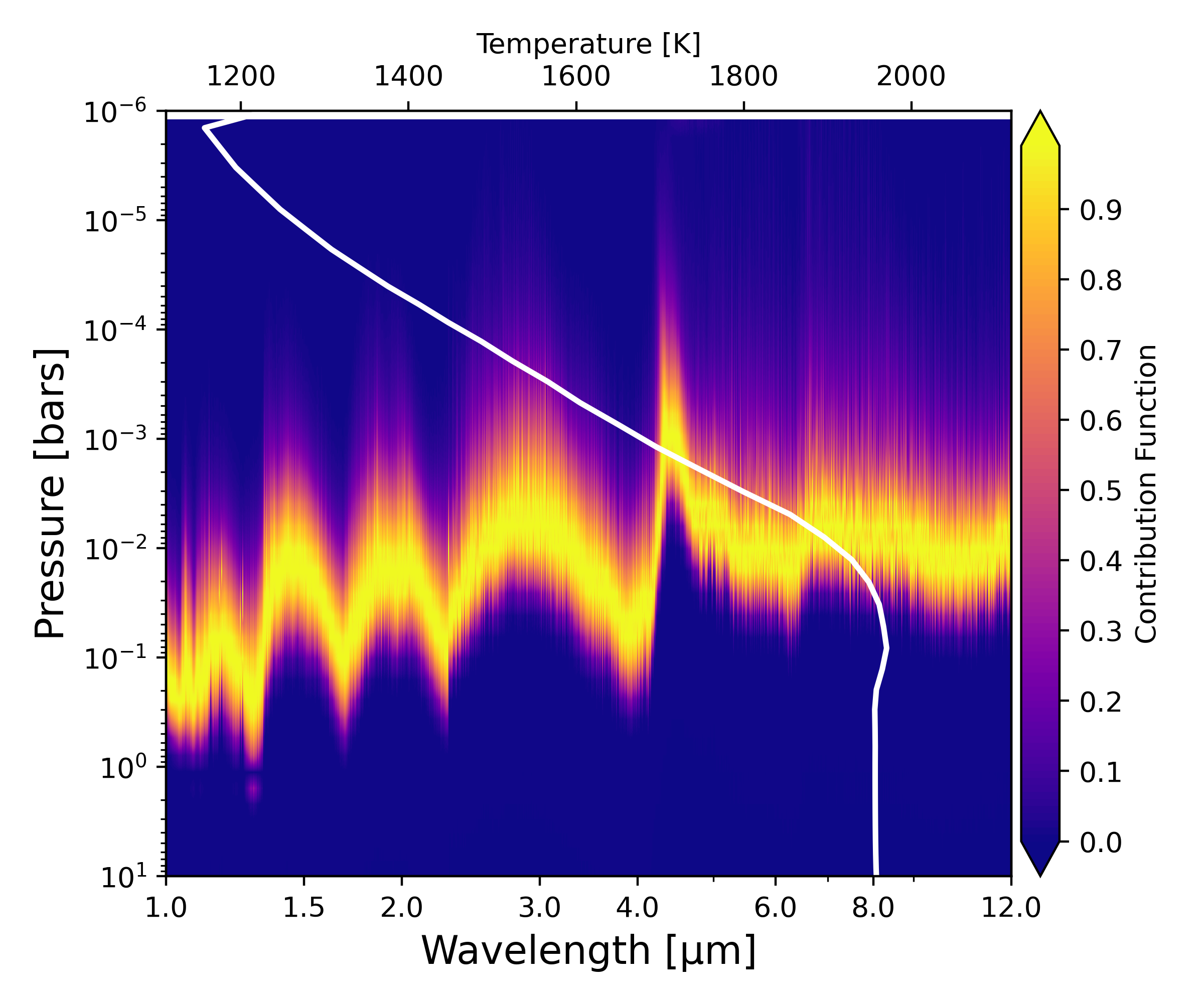}
    \caption{Figure showing contribution function for the best fit emission spectra from the solar abundance grid along with the corresponding $P$-$T$ profile (white). The color gradient ranges from yellow to blue, with yellow indicating the regions of highest contribution and blue indicating negligible contribution.}
    \label{fig:cont_func}
\end{figure}

We also compute the contribution function for our best-fit $P$-$T$ profile and the corresponding emission spectra, shown in Figure \ref{fig:cont_func}, using the methodology described in detail in \citet{Goyal2020}. We find that for the best-fit emission spectra in our work the contribution in the CO$_2$ spectral band-pass (4 to 4.2 $\mu$m) is from the pressure ranges $\sim$0.5 - 10$^{-4}$ bar. However, in \citetalias{gagnebin2024} the contribution function shows sensitivity to $\sim$0.05-10$^{-4}$ bar pressure levels in the CO$_2$ spectral band-pass. Therefore, in our work we probe deeper layers of the atmosphere in the \ce{CO2} bandpass where the temperature is cooler compared to \citetalias{gagnebin2024}. Therefore, the abundance of \ce{CO2} has to be comparatively higher to match the observed \ce{CO2} feature and this is achieved by increasing the metallicity, thus explaining the reason for our constrained metallicity being higher than that of \citetalias{gagnebin2024}.

\begin{table*}
    \centering
    
    \begin{tabular}{lcccccccc}
        \hline
        & \multicolumn{5}{c}{This work} & \cite{Bean_2023} & \multicolumn{2}{c}{\cite{gagnebin2024}} \\
        \hline
        & Solar ([O/H]) & Solar ([C/H]) & Stellar & 12 opacity & 12 opacity+VO &  & without VO & with VO \\
        \hline
        \\
        $\log Z$ & $1.94^{+0.11}_{-0.21}$ & $1.96^{+0.16}_{-0.27}$ & $1.76^{+0.12}_{-0.18}$ & $1.20^{+0.14}_{-0.05}$ & $2.17^{+0.24}_{-0.24}$ & $2.09_{-0.32}^{+0.35}$ & $1.16^{+0.26}_{-0.37}$ & $1.31^{+0.18}_{-0.23}$ \\
        \\
        C/O & $0.61^{+0.07}_{-0.14}$ & $0.60^{+0.10}_{-0.29}$ & $0.62^{+0.06}_{-0.13}$ & $0.87^{+0.02}_{-0.06}$ & $0.55^{+0.04}_{-0.06}$ & $0.84_{-0.03}^{+0.03}$& $0.673_{-0.265}^{+0.06}$ & $0.71_{-0.46}^{+0.02}$ \\
        \\
        RCF & $0.73^{+0.03}_{-0.02}$ & $0.74^{+0.03}_{-0.04}$ & $0.73^{+0.03}_{-0.03}$ & $0.92^{+0.04}_{-0.06}$ & $0.86^{+0.05}_{-0.03}$ &  & $1.17^{+0.13}_{-0.10}$ & $0.91^{+0.05}_{-0.05}$ \\
        \\
        \hline
    \end{tabular}
    \caption{Retrieved atmospheric parameters of HD-149026 b derived from this work and compared with previous studies. The table includes the metallicity ($ \log Z $), carbon-to-oxygen ratio ($ $C/O$ $), and energy recirculation factor (RCF) determined under solar and stellar abundance references. Results from this work are shown alongside the findings of \citetalias{Bean_2023} and \citetalias{gagnebin2024}. Uncertainties for each parameter are represented as the 1$\sigma$ confidence intervals.}
    \label{tab:retrieval_comparison}
    \begin{minipage}{0.9\linewidth}
    \end{minipage}
\end{table*}


\section{Discussion}\label{discussion}

\begin{figure*}
    \centering
    \includegraphics[width=\linewidth]{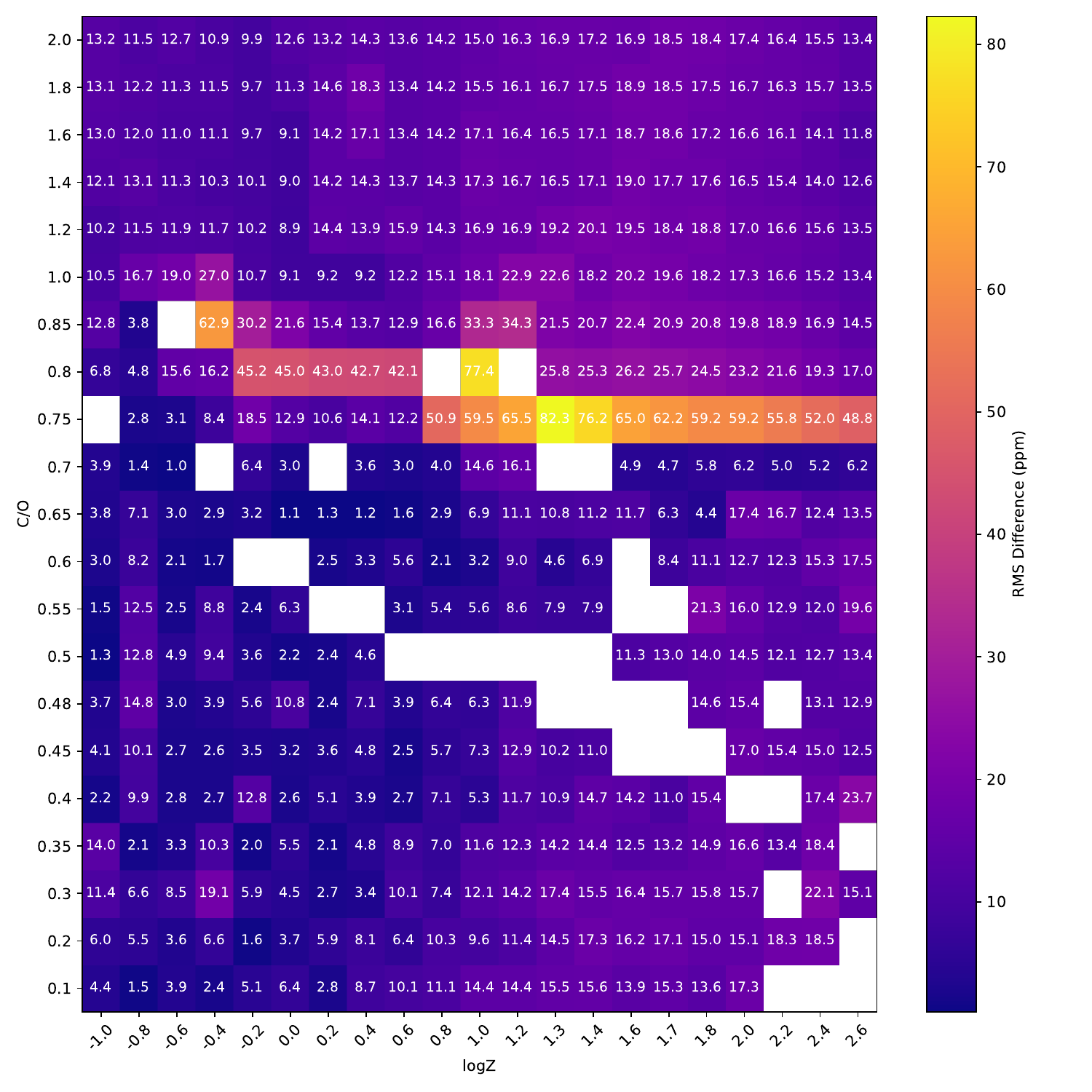}
    \caption{ Heatmap showing the root mean square (RMS) difference (in ppm) between the model emission spectra (1 to 15$\mu$m) using solar and stellar abundances across a grid of carbon-to-oxygen (C/O) ratios and metallicities (logZ). The RCF value adopted for the plot is 0.7. Each cell is annotated with the corresponding RMS difference value. White spaces represent grid points where self-consistent models did not converge. The color bar on the right shows the RMS difference, with lower values in dark purple and higher values in yellow.}
    \label{fig:rms_diff}
\end{figure*}

Constraints on C/O ratio and metallicity of exoplanet atmospheres can be very useful to constrain the formation location of exoplanets \citep{oberg2011}. However, the constraints on C/O and metallicity can depend on the choice of elemental abundances used in exoplanet atmosphere models. 
Solar elemental abundances are commonly used in exoplanet atmosphere models due to non-availability of host star elemental abundances. However, for many planetary systems the host star abundances can be quite different from that of the Sun. In this work we highlight how the choice of solar vs stellar elemental abundances can influence the constraints on the C/O ratio and metallicity of exoplanet atmospheres. The abundances of all the elements in the host star HD-149026 are considerably higher than that of the Sun as shown in Table \ref{tab:elements}. Therefore, we performed sensitivity tests on this system to assess the impact of solar versus stellar elemental abundances, in constraining the C/O ratio and metallicity of HD-149026b. 

We find that the choice of solar versus stellar abundances essentially impacts the inventory of different elements and therefore the equilibrium abundances of various atomic and molecular species that shape the $P$-$T$ structure of the exoplanet atmosphere as well as its emission spectrum. The differences in the self-consistent $P$-$T$ profiles, corresponding chemical abundances and therefore the emission spectra, with solar versus stellar abundances are different across the parameter space of C/O ratio and metallicity. These differences (root mean square difference) in the model emission spectra between solar and stellar abundances for HD-149026b for a range of C/O ratio and metallicity is shown in Figure \ref{fig:rms_diff}. It can be noticed that the differences are not substantial (< 30 ppm)for most of the C/O ratio and metallicity values. However, substantial differences going maximum upto 80 ppm can be seen for C/O values between 0.75 and 0.85. This highlights the regime where we could expect to see deviation in the inferred metallicity and C/O ratio if we choose solar abundances instead of stellar abundances to interpret emission spectra observations of HD-149026b. This regime of C/O ratio from 0.75 to 0.85 is also the regime where there is sharp transition in the abundances of carbon and oxygen bearing species, highlighted by various previous studies with solar elemental abundances \citep{2015_Molliere,Goyal_2018}. Specifically, the abundances of H$_2$O and CO$_2$ decrease substantially, while that of CH$_4$ and C$_2$H$_2$ increase, as C/O value increases from 0.7. Since this is the regime of maximum gradient and the inventory of oxygen and carbon is different between solar and stellar abundances, we see the largest difference between the emission spectra in this regime. At lower (<0.7)and higher C/O ratios (>0.85), the atmosphere is oxygen rich and oxygen poor, respectively, and they remain the same with solar and stellar abundances, therefore leading to negligible differences. Although, in this work we have performed these sensitivity tests for HD-149026 system, the results of the sensitivity tests also applies for other systems. In summary, using the host star's abundances, when available, will lead to more accurate constraints on atmospheric metallicity and C/O ratio of exoplanet atmospheres, which will also be beneficial for fair comparison between different planetary systems. 

Chemical-equilibrium retrievals and self-consistent models both contribute to our understanding of exoplanet atmospheres, but in distinct ways. Chemical-equilibrium retrievals infer atmospheric properties by assuming chemical equilibrium with parametric $P$-$T$ profiles and fitting these parameters to observational data. They reduce the number of free parameters compared to free retrievals, making the interpretation more physically motivated while still guided by observations. Self-consistent models, on the other hand, compute the atmospheric structure by solving for radiative-convective and chemical equilibrium self-consistently, without fitting to data. While chemical-equilibrium retrievals determine the best-fit equilibrium parameters that match observations, self-consistent models test whether those retrieved conditions arise naturally from first principles. Combining both methods allows us to interpret data within the framework of physically plausible atmospheric scenarios.

\citetalias{Bean_2023} presented the JWST NIRCam emission spectra observations of HD-149026b and constrained its metallicity to be 59-276 $\times$ solar, using chemical equilibrium retrievals. On the other hand, \citetalias{gagnebin2024} applied self-consistent model grid to the same observations and obtained a significantly lower metallicity of $10-20 \times$ solar. These two studies yield notably different results despite analyzing the same observational data. \citetalias{gagnebin2024} suggested that these discrepancies arise due to differences in the retrieved versus self-consistent temperature-pressure ($P$-$T$) profiles, especially in the upper atmosphere. To investigate this further, we conducted a detailed atmospheric study of HD-149026 b using the observational data from \citetalias{Bean_2023} and self-consistent models similar to \citetalias{gagnebin2024}, but with a different model and model choices. We generated five grids of model atmospheres with different choices, as described earlier in the paper. Interestingly, we obtained consistently high metallicities across four of these grids when fitting to observations and only in the grid where we include opacities similar to \citetalias{gagnebin2024} without VO we obtain similar result as that of \citetalias{gagnebin2024}. Therefore, the main underlying cause behind these differences in constrained metallicity is the choice of molecular opacities that we include in our model. We find that it is the inclusion of \ce{Fe} opacities that leads to differences in the results of this work and \citetalias{gagnebin2024}. \ce{Fe} opacity leads to comparatively cooler $P$-$T$ profile ($\sim$300\,K cooler) in the deeper atmosphere as shown in Figure \ref{Fe_effect}, and hence the abundance of \ce{CO2} needs to be increased to fit the observational feature, and that results in high obtained metallicity. We constrain the metallicity of HD-149026b to be $\sim54-112 \times$ solar metallicity with solar abundance grid, consistent with that of \citetalias{Bean_2023} but with tighter constraints. We also investigated the impact of adopting host star abundances instead of the commonly used solar abundances. We constrain the metallicity of HD-149026b to be $\sim38-76 \times$ stellar metallicity with host star abundance grid.

The atmospheric metallicity constraint that we obtain appears significantly higher than what we would expect from the mass-metallicity relationship observed in giant planets (\citet{Kreidberg_2014, Thorngren_2016}). The mass-metallicity trend suggests that lower-mass planets tend to have higher metallicities due to their formation process, where they accrete more heavy elements relative to their gas envelope. However, with a mass of 67 earth masses \citep{2005_Sato}, our HD-149026 b’s inferred metallicity of $\sim 54-112 \times$ solar is much higher than predicted by this trend, which typically suggests a metallicity of $\sim10-30 \times$ solar for a planet of this mass. This discrepancy could indicate that HD-149026 b underwent a unique formation history. One possibility is that it experienced a high degree of solid accretion, either through planetesimal bombardment or core erosion, enriching its envelope beyond typical levels \citep{Mordasini_2016}. Another explanation could involve the effects of its host star's composition, since HD-149026 is already metal-rich, the planet may have formed in an environment that naturally favored a higher-metallicity atmosphere. Understanding whether HD-149026 b is an outlier or part of a broader population of metal-rich planets will require further studies of planets in a similar mass range around metal rich stars. Future observations, particularly of atmospheric composition across different planetary masses, will help refine our understanding of the mass-metallicity relationship and whether additional processes contribute to the observed enrichment. More observations, particularly with JWST, can improve our understanding by providing better constraints on key molecules and refining metallicity and C/O estimates. Future observations, combined with improved modeling techniques, will be essential for obtaining a clearer picture of HD-149026 b’s atmospheric composition and formation history.

\section{Conclusions} \label{Conclusion}

In this work we highlighted the differences in the self-consistent $P$-$T$ profiles, chemical abundances and thereby the emission spectrum, due to choice of solar versus stellar elemental abundances in the exoplanet atmosphere models. We further interpreted the JWST NIRCam emission spectra observations of HD-149026b using grid of self-consistent model atmospheres, with different model choices. We created five grids of self-consistent model atmospheres; first with solar elemental abundances and varying [O/H] to vary $C/O$ ratio, second with stellar elemental abundances and varying [O/H] to vary $C/O$ ratio, third with solar elemental abundances and varying [C/H] to vary $C/O$ ratio, fourth with opacities similar to that adopted in \citet{gagnebin2024} without VO and fifth with opacities similar to that adopted in \citet{gagnebin2024} including VO. We finally compared the constraints that we obtain with these different grids with the constraints obtained by \citet{gagnebin2024} and 
\citet{Bean_2023} for HD-149026b. The major conclusions from our work are as follows:

\begin{itemize}

    \item Using solar instead of stellar elemental abundances in the models to interpret the observations of exoplanet atmospheres, can lead to differences in the constrained metallicity and C/O ratio in certain range of parameter values. For HD-149026b we find the difference in the emission spectra to be maximum upto 80 ppm, for C/O ratios between 0.75-0.85.

    \item Using self-consistent atmosphere model grid with solar abundances, we constrain the metallicity of HD-149026b to be 53–113 $\times$ solar metallicity. This is higher than the previous constrained range of 12-31 $\times$ solar metallicity by \citet{gagnebin2024}. However, our constraints are close to the metallicity range of 59-276 $\times$ solar metallicity constrained by \citet{Bean_2023}, using chemical equilibrium retrievals.

    \item Using self-consistent atmosphere model grid with stellar abundances, we constrain the metallicity of HD-149026b to be 38–76 $\times$ stellar metallicity. This indicates that the planet atmosphere is enriched in heavier elements relative to its host star. 

    \item  We constrain the $C/O$ ratio of HD-149026b between 0.46 to 0.68 using both solar and stellar abundance grid. The retrieved C/O ratio using solar and stellar abundances are similar, as it lies outside the range of 0.75-0.85 where maximum differences could be seen. Our C/O ratios are consistent with the C/O values reported by \citet{gagnebin2024} but significantly less than $0.84^{+0.03}_{-0.03}$ reported by \citet{Bean_2023}.

    \item We constrain the self-consistent $P$-$T$ profile of HD-149026b to be substantially cooler (upto 500\,K)  than \citet{gagnebin2024} in the emission spectra probed region (0.0001 to 1 bar), thus requiring higher abundance of CO$_2$ to explain the observed JWST NIRCam features, leading to comparatively higher metallicity constraint. 
    
    \item We find that inclusion of Fe opacity in computing self-consistent $P$-$T$ profiles for HD-149026b in our models is the major reason for the differences in the metallicity constraints between this work and \citet{gagnebin2024}.

    \item We constrain the recirculation factor of HD-149026b to be between 0.70 and 0.76 with both solar and stellar abundance grids. This is considerably lower than the values reported by \citet{gagnebin2024}, indicating higher energy transport to the night-side.

\end{itemize}
Our current analysis relies on 1D atmospheric models, which assume a horizontally uniform structure and capture few 3D effects (recirculation factor) in oversimplified ways. Given the highly irradiated nature of HD-149026 b, these 3D processes could significantly impact the thermal structure and chemical abundances, influencing the observed spectra. Additionally, our constraints are limited by the specific wavelength range of the available observations, which may not fully capture key molecular features or thermal signatures. Moreover, in the current study we did not include any dis-equilibrium processes like vertical mixing and photochemistry in our model. A more comprehensive approach incorporating 3D effects, dis-equilibrium processes and extending observations across a broader wavelength range would be necessary to refine our understanding of the atmospheric properties of HD-149026 b.

\section*{Acknowledgements}
We thank the anonymous referee for their constructive suggestions that has improved the clarity of this manuscript. JG acknowledges funding from the SERB SRG Grant SRG/2022/000727-G of the Department of Science and Technology (DST), Government of India, that provided computational facilities for this work. We thank Avinash Verma and Swaroop Avarsekar for their valuable feedback and time, which helped improve the clarity and readability of the manuscript. We extend our thanks to Avinash Verma for producing Fe opacities in the extended wavelength range to perform some sensitivity tests.


\section*{Data Availability}
The JWST observations used in this work are publicly available. The model data underlying this article will be shared on reasonable request to the corresponding author.



\bibliographystyle{mnras}






\appendix

\section{}
\bsp	


\begin{figure}
    \centering
    \includegraphics[width=\linewidth]{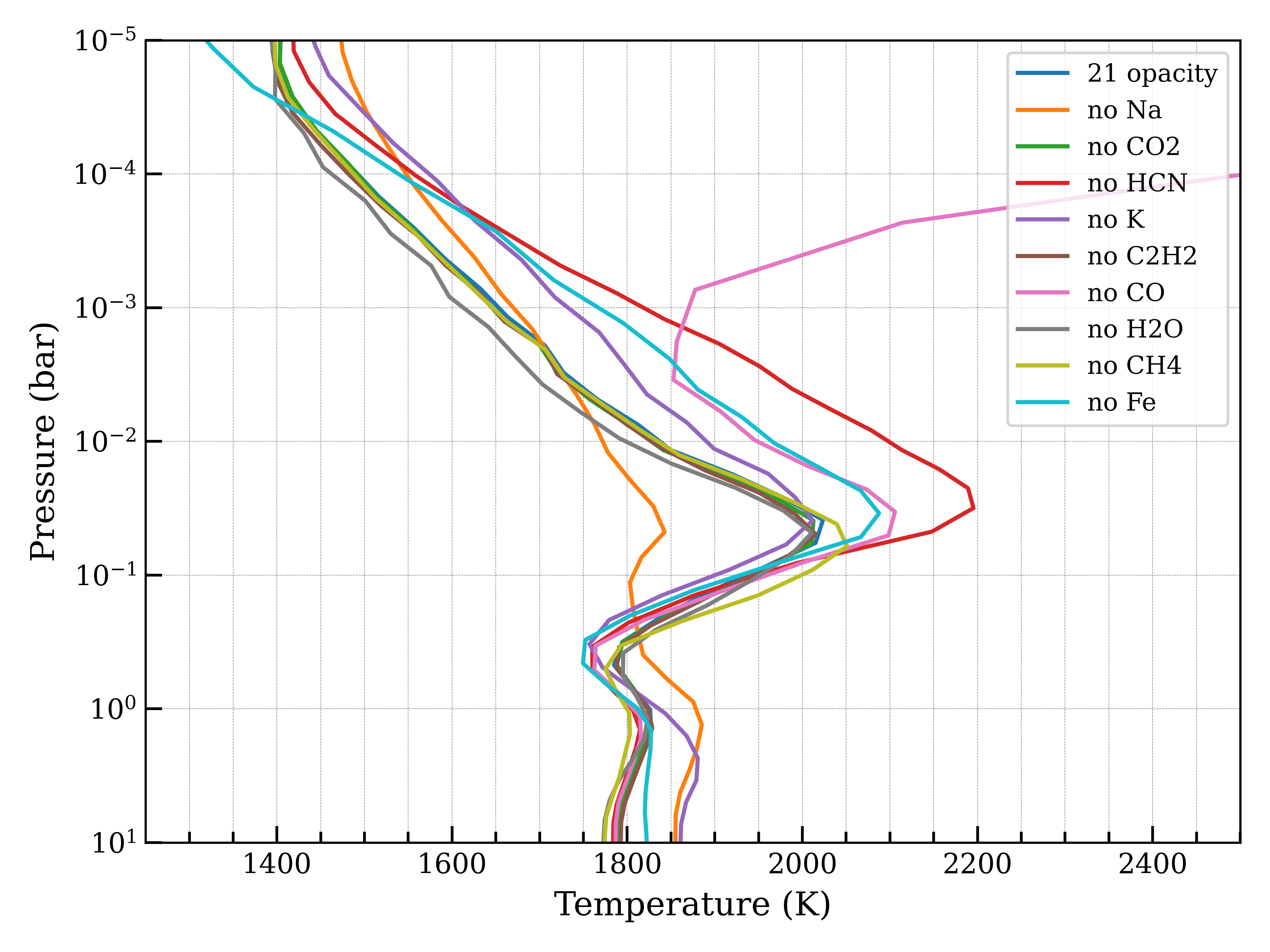}
    \caption{Pressure-Temperature profile showing the contribution of molecules to the temperature inversion for $C/O$ = 0.8 in Figure \ref{fig:Sens CO PT}. }
    \label{fig:inversion_molecule_PT_sens_CtoO}
\end{figure}

\begin{figure}
    \centering
    \includegraphics[width=\linewidth]{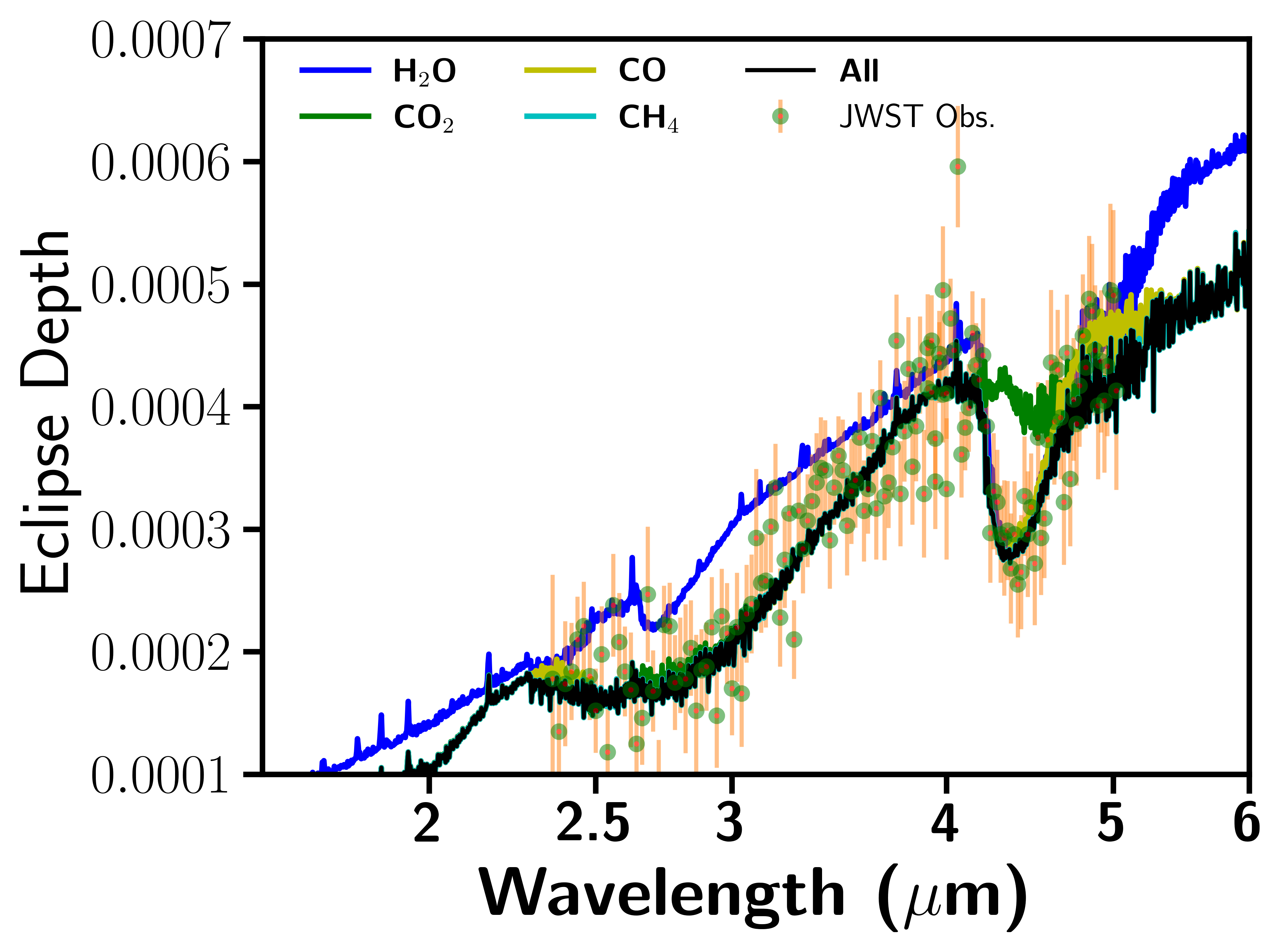}
    \caption{Figure showing emission spectra features of individual
molecule used in ATMO for best-fit emission spectra using solar abundances. \ce{H2O} (blue), \ce{CO2} (green), \ce{CO} (olive), \ce{CH4} (cyan), and all 20 opacities (black). The simulation with just (\ce{CO}) in olive shows a few spectral features at 2.4 microns and 5 microns. Comparing the all opacity simulation in black, and the simulation in green with just \ce{CO2} opacity leads to a spectral feature 4-5 $\mu$m. Simulation with only \ce{H2O} in blue leads to spectral features in all the wavelength ranges except in the range of 4-5 microns}.
    \label{fig:opacity}
\end{figure}

\begin{figure}
    \centering
    \includegraphics[width=\linewidth]{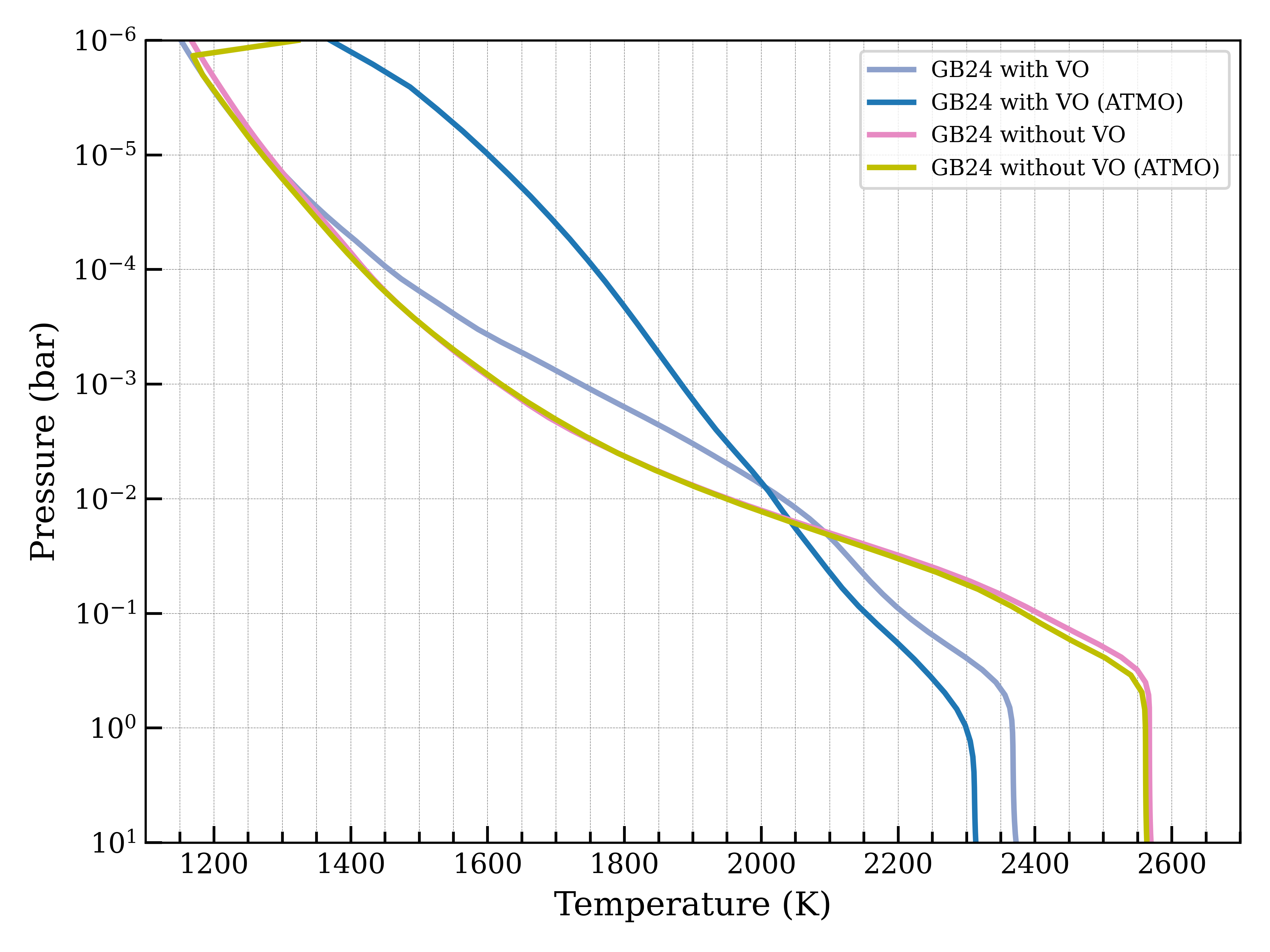}
    \caption{Temperature-pressure (T-P) profile from \citetalias{gagnebin2024}, regenerated using ATMO with their best-fit results and the opacities listed in the paper.}
    \label{fig: no VO gagnebin PT}
\end{figure}

 \begin{figure}
    \centering
    \includegraphics[width=\linewidth]{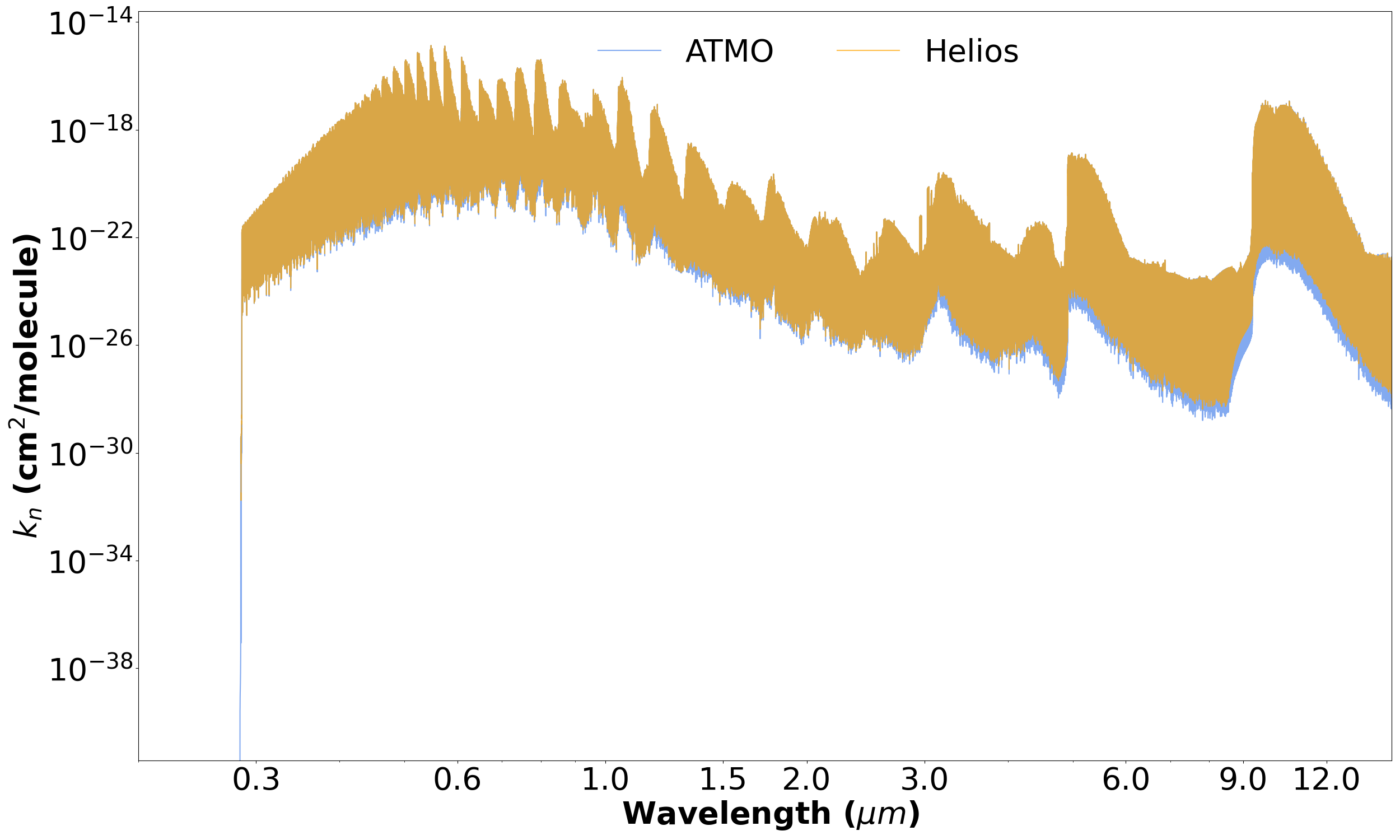}
    \caption{Comparison of VO absorption cross-section at a pressure of 1mbar and 1500 K temperature obtained from Helios-K (orange) and that in ATMO (blue).}
    \label{fig: VO abs}
\end{figure}

\begin{figure}
     \centering
    \includegraphics[width=\linewidth]{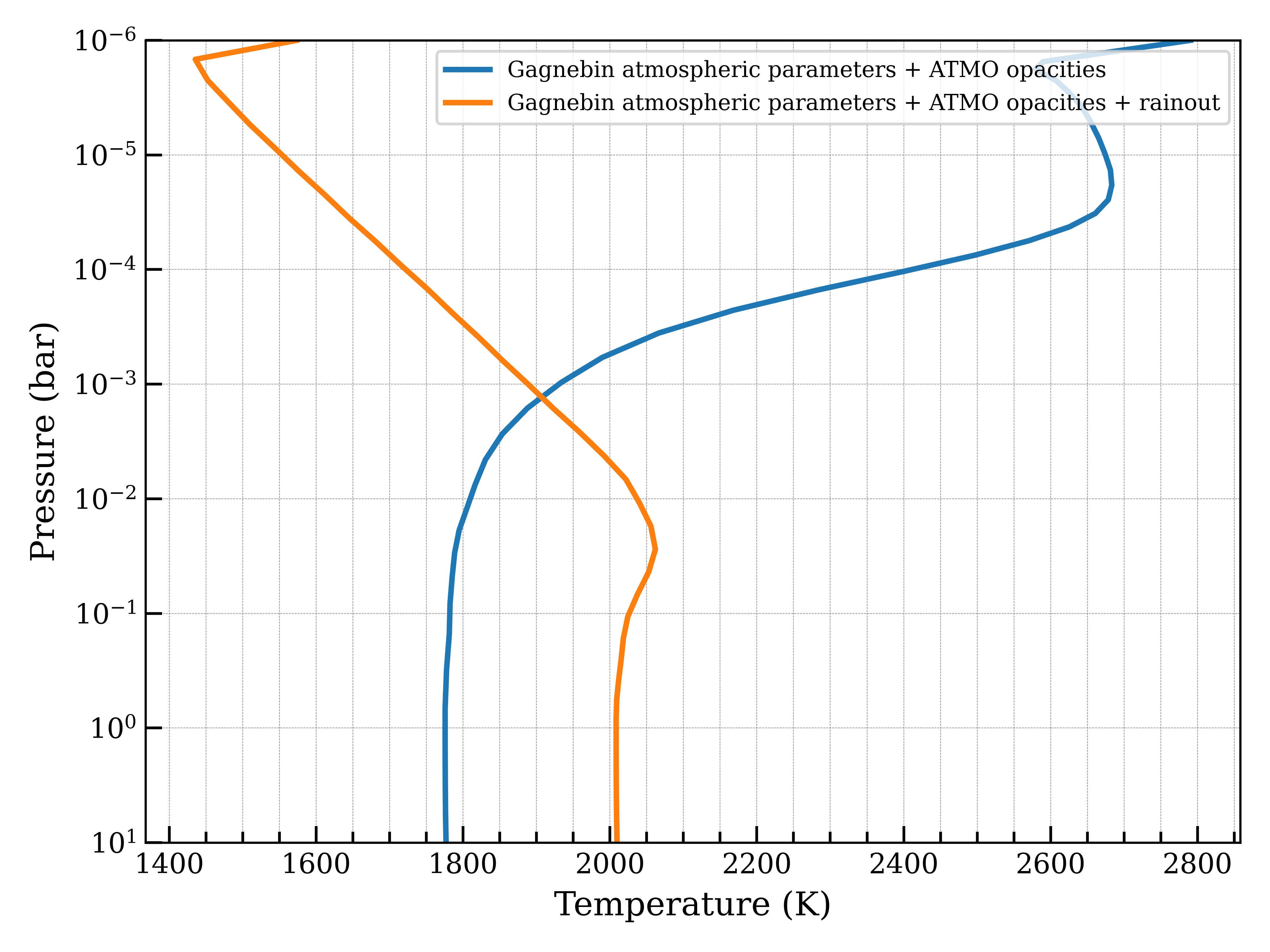}
    \caption{Pressure-temperature ($P$-$T$) profiles illustrating the effect of rainout condensation, generated using ATMO. The profiles are computed based on the best-fit median parameters from \citetalias{gagnebin2024}, incorporating the opacity contributions of 21 molecular species.}
    \label{fig:rainout}
\end{figure}

\begin{figure}
    \centering
    \includegraphics[width=0.9\linewidth]{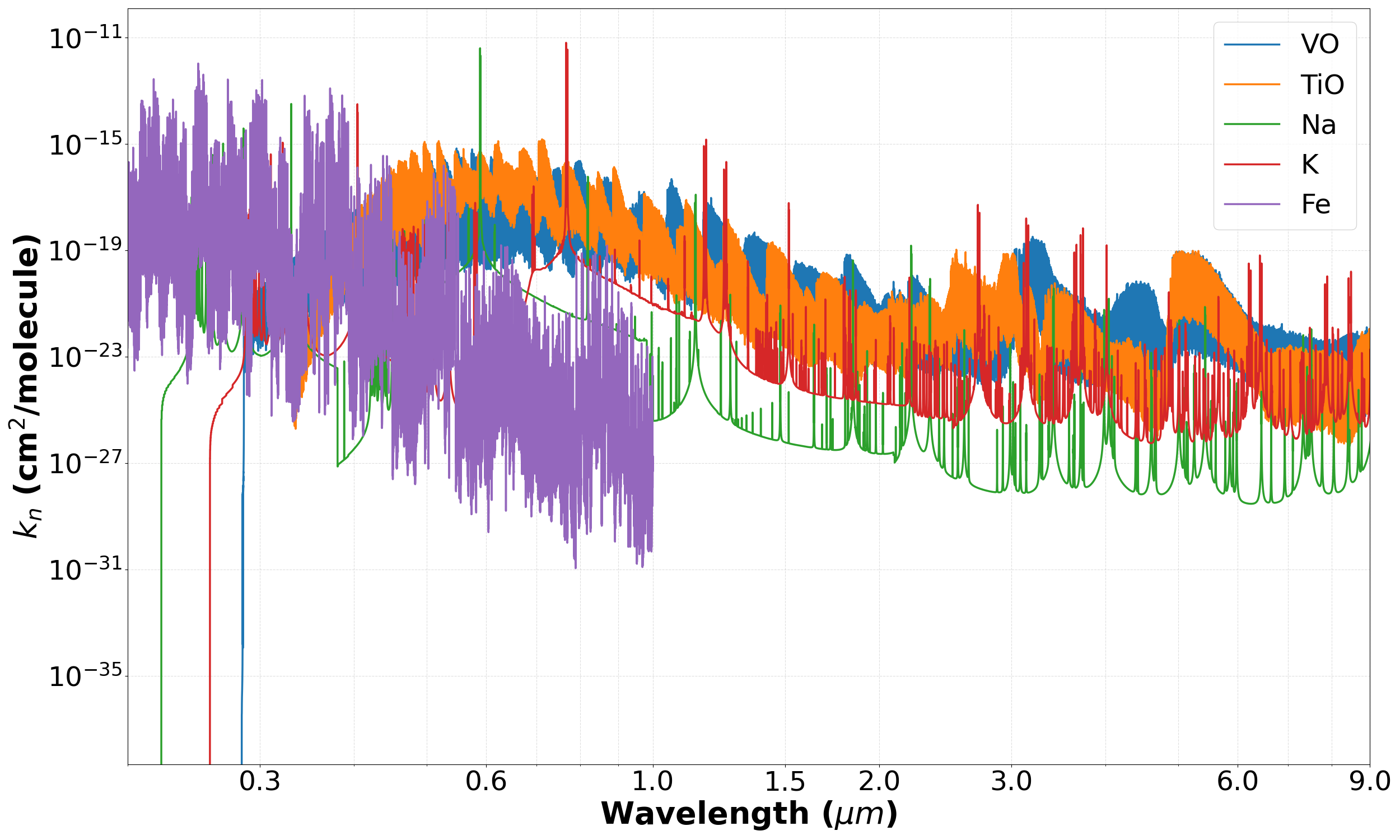}
    \caption{ Absorption cross-sections of Fe, TiO, VO, Na, and K computed at representative atmospheric conditions (T = 2100 K, P = 0.1 bar). The cross-section of Fe is comparable in magnitude to those of other strong optical absorbers, suggesting that under certain conditions, Fe may contribute to the overall opacity and potentially influence the thermal structure of the atmosphere.
}
    \label{fig:xsec}
\end{figure}



\label{lastpage}
    \end{document}